\documentclass[journal]{IEEEtran}
\usepackage{cite}
\ifCLASSINFOpdf
   \usepackage[pdftex]{graphicx}
\else
\fi
\usepackage{amsmath}
\usepackage{mathtools}
\usepackage{amssymb}
\usepackage[monochrome]{xcolor}
\usepackage[nolist,nohyperlinks]{acronym}

\usepackage{algorithm}
\usepackage[noend]{algpseudocode}

\usepackage{stfloats}
\usepackage{url}
\usepackage[labelformat=simple]{subcaption}
\renewcommand{\Re}[1]{\mathfrak{Re}\left\{#1\right\}}
\renewcommand{\Im}[1]{\mathfrak{Im}\left\{#1\right\}}

\acrodef{UE}{user equipment}
\acrodef{BS}{base station}
\acrodef{Tx}{transmitter}
\acrodef{Rx}{receiver}
\acrodef{CFO}{carrier frequency offset}
\acrodef{ULA}{uniform linear array}
\acrodef{OFDM}{orthogonal frequency division multiplexing}
\acrodef{AWGN}{additive white Gaussian noise}
\acrodef{LO}{local oscillator}
\acrodef{MAP}{maximum \textit{a posteriori}}
\acrodef{mmWave}{millimeter wave}
\acrodef{subTHz}{sub-terahertz}
\acrodef{LOS}{line-of-sight}
\acrodef{NLOS}{non-line-of-sight}
\acrodef{MCMC}{Markov chain Monte Carlo}
\acrodef{VB}{variational Bayes}
\acrodef{EP}{expectation propagation}
\acrodef{KL}{Kullback-Leibler}
\acrodef{SBL}{sparse Bayesian learning}
\acrodef{RV}{random variable}
\acrodef{PDF}{probability density function}
\acrodef{EM}{expectation-maximization}
\acrodef{5G}{fifth generation}
\acrodef{iid}{independent identically distributed}
\acrodef{ML}{maximum likelihood}
\acrodef{SCA}{successive convex approximation}
\acrodef{MM}{majorization-minimization}
\acrodef{ADMM}{alternating direction method of multipliers}
\acrodef{SNR}{signal-to-noise ratio}
\acrodef{AoA}{angle of arrival}
\acrodef{AoD}{angle of departure}
\acrodef{MSE}{mean squared error}
\acrodef{RMSE}{root mean squared error}
\acrodef{ReMSE}{relative mean squared error}
\acrodef{PSO}{particle swarm optimization}
\acrodef{NM}{Nelder-Mead}
\acrodef{B5G}{beyond 5G}
\acrodef{6G}{sixth generation standard}
\acrodef{MIMO}{multiple input multiple output}
\acrodef{LLN}{law of large numbers}
\acrodef{LGR}{large grid regime}
\acrodef{LPA}{long pilot approximation}
\acrodef{MC}{Monte Carlo}
\acrodef{ToF}{time of flight}
\acrodefplural{TOF}[TOFs]{times of flight}
\acrodef{CACC}{cross-antenna cross-correlation}
\acrodef{MMSE}{minimum mean square error}
\acrodef{DTFT}{discrete-time Fourier transform}
\acrodef{DFT}{discrete Fourier transform}
\acrodef{FFT}{fast Fourier transform}
\acrodef{2P2T}{two precoders two transmissions}
\acrodef{TVP}{time-varying precoder}
\acrodef{RF}{radio frequency}
\acrodef{PSF}{point spread function}
\acrodef{DBSCAN}{density-based spatial clustering of applications with noise}
\acrodef{SSE}{Sum Square Error}
\acrodef{TDMA}{time division multiple access}
\acrodef{FDMA}{frequency division multiple access}
\acrodef{CDMA}{code division multiple access}
\acrodef{BER}{bit error rate}
\acrodef{MUSIC}{multiple signal classification}
\acrodef{URA}{uniform rectangular array}
\acrodef{JCS}{joint communication and sensing}
\acrodef{NPNT}{N-precoders N-transmissions}
\acrodef{SOR}{successive over-relaxation}
\acrodef{COMPAS}{concurrent mapping, positioning, and synchronization}
\acrodef{SLAM}{simultaneous localization and mapping}
\acrodef{AECD}{alternating exact coordinate descent}
\acrodef{ISAC}{integrated sensing and communications}
\acrodef{JSC}{joint sensing and communication}
\acrodef{FR2}{frequency range 2}
\acrodef{CSI}{channel state information}
\acrodef{SAGE}{space-alternating generalized expectation-maximization}
\acrodef{OR}{over-relaxation}
\acrodef{AIC}{Akaike information criterion}
\acrodef{SIMD}{single instruction multiple data}
\acrodef{ASIC}{application specific integrated circuit}
\acrodef{FPGA}{field-programmable gate array}
\acrodef{MAC}{multiply–accumulate}
\acrodef{CP}{canonical polyadic}
\acrodef{ESPRIT}{estimation of signal parameters via rotational invariant techniques}
\acrodef{MHR}{multidimensional harmonic retrieval}
\acrodef{i.i.d.}{independent identically distributed}
\acrodef{HST}{high-speed train}
\acrodef{ICI}{intercarrier interference}
\acrodef{DFRC}{dual-functional radar communications}
\acrodef{LS}{least squares}
\acrodef{MSVD}{multilinear singular value decomposition}
\acrodef{ALS}{alternating least squares}
\acrodef{dGN}{damped Gauss-Newton}
\acrodef{WSNSAP}{wideband spatial nonstationary wireless channels with antenna polarization}
\acrodef{GN}{Gauss-Newton}
\acrodef{RiMAX}{Richter's maximum likelihood estimation}
\acrodef{BIC}{Bayesian information criteria}
\acrodef{XL-MIMO}{extra-large MIMO}
\acrodef{FR2}{frequency range 2}
\acrodef{FR3}{frequency range 3}
\acrodef{MDL}{minimum description length}
\acrodef{mMIMO}{massive MIMO}
\acrodef{NLS}{nonlinear least squares}
\acrodef{MAE}{mean absolute error}
\acrodef{MedAE}{median absolute error}
\acrodef{CDF}{cumulative distribution function}
\acrodef{SIC}{successive interference cancellation}
\acrodef{MC}{Monte Carlo}
\acrodef{QPSK}{quadrature phase-shift keying}
\acrodef{std-SAGE}{standard SAGE}
\acrodef{PCE}{parametric channel estimation}
\acrodef{VB}{variational Bayes}
\acrodef{ARD}{automatic relevance determination}
\acrodef{IARD}{incremental automatic relevance determination}
\acrodef{PFA}{probability of false alarm}
\acrodef{LLR}{log-likelihood ratio}
\acrodef{GLRT}{generalized likelihood ratio test}
\acrodef{DOF}{degrees of freedom}

\acrodef{PCE}{parametric channel estimation}
\acrodef{ADC}{analog-to-digital converter}
\acrodef{LNA}{low noise amplifier}
\acrodef{LM}{Lloyd-Max}
\acrodef{DAC}{digital-to-analog converter}
\acrodef{PA}{power amplifier}
\acrodef{CLT}{central limit theorem}
\acrodef{ZC}{Zadoff-Chu}
\acrodef{PAPR}{peak-to-average power ratio}
\acrodef{MM}{majorization-minimization}
\acrodef{4G}{fourth generation}
\acrodef{LTE}{long term evolution}
\acrodef{CS}{compressive sensing}
\acrodef{HOSVD}{higher order singular value decomposition}
\acrodef{MS}{multiple start}
\acrodef{DPD}{digital pre-distortion}
\acrodef{EVM}{error vector magnitude}
\acrodef{PSD}{power spectral density}
\acrodef{IIR}{infinite impulse response}
\acrodef{SSB}{single sideband}
\acrodef{ZZLB}{Ziv-Zakai lower bound}
\acrodef{WWLB}{Weiss-Weinstein lower bound}
\acrodef{CRLB}{Cramér-Rao lower bound }
\acrodef{IQ}{in-phase quadrature}
\acrodef{SOTA}{state-of-the-art}
\acrodef{AER}{average error reduction}
\acrodef{ANPE}{average normalized power efficiency}

\acrodef{PPE}{path parametric error}
\acrodef{BD}{Bussgang decomposition}

\usepackage{hyperref}

\usepackage{mathabx}

\usepackage{diagbox}

\usepackage{multicol}
\usepackage{mathtools}

\usepackage{amsthm}

\usepackage{amsmath}               
  {
      \theoremstyle{plain}
      \newtheorem{assumption}{Assumption}
  }

\usepackage{xpatch}
\makeatletter
\AtBeginDocument{\xpatchcmd{\@thm}{\thm@headpunct{.}}{\thm@headpunct{}}{}{}}
\makeatother

\newcommand{\pushcode}[1][1]{\hskip\dimexpr#1\algorithmicindent\relax}

\usepackage{placeins}

\begin{document}
\title{Parametric Channel Estimation with Hardware Impaired Hybrid Beamformers: Sensing, Communications, and Power Efficiency Tradeoffs}


\author{Enrique~T.~R.~Pinto,~\IEEEmembership{Graduate Student Member,~IEEE}, 
        Silvio~Mandelli,~\IEEEmembership{Member,~IEEE},
        Marcus~Henninger,~\IEEEmembership{Member,~IEEE},
        and~Markku~Juntti,~\IEEEmembership{Fellow,~IEEE}% <-this % stops a space
%\thanks{M. Shell was with the Department
%of Electrical and Computer Engineering, Georgia Institute of Technology, Atlanta,
\thanks{Enrique T. R. Pinto and Markku Juntti are with the Centre for Wireless Communications, University of Oulu, 90014 Oulu, Finland (e-mail:\{enrique.pinto, markku.juntti\}@oulu.fi). 
Marcus~Henninger and \mbox{Silvio}~Mandelli are with Nokia Bell Labs Stuttgart, 70435 Stuttgart, Germany (e-mail:\{marcus.henninger, silvio.mandelli\}@nokia-bell-labs.com). The work was supported in part by the Research Council of Finland (former Academy of Finland) 6G Flagship Program (Grant Number: 369116), 6GWiCE project (357719), and S6GRAN project (370559).}% <-this % stops a space
}


\markboth{}%Journal of \LaTeX\ Class Files,~Vol.~14, No.~8, August~2015}%
{}%Shell \MakeLowercase{\textit{et al.}}: Bare Demo of IEEEtran.cls for IEEE Journals}



\maketitle

\begin{abstract}

Due to high power consumption and hardware costs of fully digital arrays, hybrid beamformers are often considered as a more economic alternative. Furthermore, using high resolution analog to digital converters (ADCs) can also have prohibitive power consumption, which leads to lower resolution converters being considered for radio frequency (RF) front end design. The finite quantization resolution as well as the nonlinearities caused by the power amplifiers (PAs) and low noise amplifiers (LNAs) can have a substantial impact on system performance. While widely studied for communications, the impact of hardware impairments on sensing performance is considerably less explored. In this work, we study the interplay between hybrid beamforming architectures, hardware impairments, and sensing and communications performance. Additionally, we define the concept of double-isotropy for pilot-combiner pairs, formalizing the notion of a perfectly energy-fair beam sweep. The multiple start (MS) space-alternating generalized expectation-maximization algorithm (SAGE) is also introduced, aimed at addressing the optimization issues arising from parametric channel estimation (PCE) in hybrid beamformed systems. We then provide a set of numerical results assessing the impacts of beamformer architecture and ADC resolution on PCE, sensing, and communications performance. The results show that medium resolution ADCs lead to the most power efficient configurations, with the best tradeoff between power consumption and performance for the majority of beamforming architectures. Additionally, fully digital beamforming architectures with high resolution converters can often be substituted for a hybrid beamformer setup with medium resolution converters without significant performance loss at a lower power consumption and overall hardware cost.
%The results show that fully digital beamforming architectures with high resolution converters can often be substituted for a hybrid beamformer setup with medium resolution converters without significant performance loss and at a lower power consumption and overall hardware cost. Furthermore, fully analog arrays with high resolution converters are not robust to hardware impairments and display poor overall performance within the investigated conditions.

\end{abstract}

% with a focus on integrated sensing and communications (ISAC) in the upper-midband,


% \fi
\IEEEpeerreviewmaketitle


%\IEEEpubidadjcol

\section{Introduction}

\IEEEPARstart{W}{ith} the current growing interest in sensing and localization as a commercial use case for \ac{B5G} and \ac{6G} systems, enabled by current advances in \ac{MIMO} systems, \ac{ISAC} has become an extremely active research area \cite{zhang_enabling_jsc,isac1,isac2, isac_imaging}. Using communications infrastructure to provide sensing services is considered to be an economically effective approach towards satisfying the existing commercial demand. Enhancing existing communications hardware with sensing capabilities is a safer way -- from a business perspective -- of satisfying potential demands without requiring major modifications to infrastructure. Thus, communications-centric approaches are currently highly preferred by the telecommunications industry~\cite{6gisac_poc}.    

%, but considered one of the only viable forms of achieving successful \ac{ISAC} within the narrow profit margins of the highly competitive telecommunications industry. % (citation needed?).\par

Simultaneously, in an effort to satisfy the increasing demand for throughput, mobile communications standards are set to occupy higher frequency bands when compared to the \ac{4G} \ac{LTE} standard, such as \ac{FR2} in the \ac{5G} standard \cite{mmwave_comm_survey} or even \ac{FR3} \cite{emil_midband, cov_vs_sensing} also known as the upper-midband. These newly available frequencies not only offer additional spectrum, but also allow the usage of reduced size antenna elements, being a crucial enabler for \ac{mMIMO} and \ac{XL-MIMO}. Larger arrays improve the spatial multiplexing capabilities of mobile systems, improving the multiuser channel capacity compared to standard \ac{MIMO} systems. On the communications side, this makes it possible to manage a higher number of simultaneous users while satisfying minimum throughput requirement. From a sensing and localization perspective, the sharper angular resolution of a larger array can reliably resolve sensing targets that are closer together \cite{overview_jcrs, rcc_overview, jrcd}. %Other benefits of \ac{mMIMO} include higher beamforming gain \cite{massivemimobook} and robustness to channel conditions through the channel hardening effect \cite{ch_hardening1, ch_hardening2}.\par

% This is copied almost straight from the spawc paper
The demand for sensing and positioning information has, in turn, led to an increased interest towards \ac{PCE} algorithms, which can be used both for sensing and communications applications. The \ac{PCE} approach differs from traditional channel estimation techniques (which recover exclusively the channel matrix or channel tensor) by exploiting sparsity to decompose the channel into individual multipath components with their own parameters. The parameters of interest typically include \ac{AoA}, \ac{AoD}, delay, and Doppler shift. This kind of processing was, for the most part, limited to the context of channel measurement campaigns, in which the objective is the extraction of path statistics with more time available for data processing. Most \ac{PCE} algorithms rely on one of the following approaches: \ac {CS} techniques such as \ac{SBL} \cite{sbl1,sbl2,sbl3}, tensor decompositions like the \ac{CP} decomposition \cite{henk_cp} and \ac{HOSVD} \cite{tensors}, or \ac{ML} estimation such as the \ac{SAGE} \cite{sage, vb_sage, ch_est_pinto} and the RiMAX \cite{rimax} algorithms. The development of more efficient \ac{PCE} algorithms is an essential step towards achieving practical \ac{ISAC} \cite{zhang_enabling_jsc} as they are the core data acquisition tool for sensing and localization algorithms.\par

Despite the wide range of performance advantages of increasing the number of antennas in the radio transceivers, there are substantial drawbacks regarding energy consumption, power efficiency \cite{energy_efficient_mmimo}, as well as the actual hardware costs, which increase the total cost of ownership of the devices. Although equipping every antenna element with a dedicated \ac{RF} chain would be ideal in terms of maximizing performance, the resulting energy demand would be too high for typical commercial systems such as \acp{BS} and \acp{UE} \cite{fd_is_expensive1,fd_is_expensive2}. A popular proposed solution for this issue is to decrease the number of \ac{RF} chains by using hybrid beamforming architectures \cite{hybrid_precoding}. With hybrid beamformers, a smaller number of \ac{RF} chains is used by performing a combination of digital and analog beamforming. This approach leads to a more flexible design process with a tradeoff between performance and energy consumption. Despite the added flexibility, the design of beamforming schemes for hybrid architectures is considerably more complicated than for fully digital systems. While digital beamformers can often be computed optimally using convex optimization methods, hybrid beamformers often rely on heuristics to avoid expensive discrete and nonconvex optimization techniques. The energy efficiency of hybrid beamformer topologies are most often studied in the context of spectral efficiency \cite{energy_eff_hybrid_spectral_eff, energy_eff_hybrid_spectral_eff_ris} with sensing aspects being, so far, mostly neglected.\par

The transceiver power consumption on the radio front-end is also related to the resolution of \acp{ADC} and \acp{DAC}. Since \ac{ADC} power consumption grows exponentially with the number of quantization bits \cite{adc_power_exp}, low-resolution \acp{ADC} are taken as a viable design choice for decreasing power consumption and potentially improving the power efficiency of mobile radio systems while still retaining an acceptable performance level \cite{onebit_adc1, onebit_adc2, onebit_adc3}. Particularly, 1-bit \acp{ADC} have received substantial attention from the wireless communications research community, with an emphasis on communications performance \cite{onebit_adc2}. Heavily quantized communication systems are well studied from the channel capacity perspective, e.g., in \cite{onebit_adc1} for 1-bit \acp{ADC} with \ac{Tx} \ac{CSI}. However, the sensing and \ac{PCE} performances have not been thoroughly explored with low-resolution converters. \par

% Paragraph on Bussgang decomposition: 
% main points: 1.works commonly assume perfect CSI tx rx (check long list of citations on page 2 end of column 1 of "Digital and Hybrid Precoding Designs in Massive MIMO With Low-Resolution ADCs")
% 2.Using the bussgang decomposition to derive more specific results requires an increasing number of assumptions on the input and output signals of the nonlinearity, i.e., loss of generality
% 3. Not practical to derive an expression for the distribution of the bussgang error, also the bussgang errors in MIMO are highly correlated for low-resolution systems
 
% References: https://ieeexplore.ieee.org/stamp/stamp.jsp?tp=&arnumber=9307295, https://oulurepo.oulu.fi/bitstream/handle/10024/56517/nbnfioulu-202505283998.pdf?sequence=1&isAllowed=y, https://ieeexplore.ieee.org/stamp/stamp.jsp?tp=&arnumber=11006401

For a theoretical treatment of nonlinear hardware impairments (such as signal quantization and amplifier nonlinearity), the \ac{BD} is often used \cite{bussgang}. The \ac{BD} decouples a nonlinearly affected signal into the input signal scaled by a gain (called the Bussgang gain) and an uncorrelated noise component (which we call the Bussgang noise). Due to the difficulty of independently estimating the channel and the \ac{Tx} and \ac{Rx} impairments, channel estimation algorithms and precoder design procedures commonly assume perfect \ac{CSI} \cite{hyb_perf_csi1,hyb_perf_csi2,hyb_perf_csi3, hyb_bf_ch_est_perf_csi}. Furthermore, using the Bussgang decomposition for \ac{MIMO} systems with correlated inputs leads to challenging computations which are simplified by forcing some assumptions on the properties of the output and input signals \cite{bussgang_assumptions_hyb_bf_adc,bussgang_diag_approx}. The validity of such assumptions is often weakly justified or unclear \cite{bussgang_assumptions_hyb_bf_adc} and leads to loss of generality. Deriving expressions for the Bussgang gain and the distribution of the Bussgang noise is also not trivial, which limits the usefulness of the \ac{BD} in estimation contexts\footnote{The Bussgang decomposition remains a very powerful tool for deriving rate expressions for more specific scenarios.}. \par

% Also hybrid beamforming channel estimation papers:
% 1. https://ieeexplore.ieee.org/document/9563221 | Assumes correlation matrix knowledge (thus channel should also be 2nd order stationary)
% 2. https://ieeexplore.ieee.org/document/8744577 | requires a 2-stage approach, shows a noticeable performance gap of HY beamformers with respect to the case of FD beamforming

Within the presented context, this work aims to provide an initial study of the interplay between beamformer architectures, hardware impairments, and the overall system performance regarding channel estimation, sensing, communications, and power efficiency. Ideally, the properties of each architecture should be studied using statistical performance bounds. However, the usual approach using the \ac{CRLB} is not appropriate for this study due to its poor characterization of the \textit{threshold} (low SNR) and the \textit{a priori} (very low SNR) regions, which are essential in our analysis. On the other hand, tighter and nonlocal bounds such as the \ac{ZZLB} \cite{ziv_zakai} or the \ac{WWLB} \cite{ww_bound} are generally challenging to compute. Thus, in this work, we use a newly derived \ac{SOTA} channel estimator as a proxy to the best achievable performance in each scenario.

The design of the channel estimation algorithm also hints towards a desirable property for pilots and combiners: invariance of the total received and delivered energy with respect to \ac{AoA} and \ac{AoD} within a pilot frame. This is a conceptual extension of the pilot isotropy condition defined in \cite{ch_est_pinto} to a hybrid beamforming setup. A pilot-combiner pair that satisfies this property is called a \textit{doubly isotropic} pilot-combiner pair. Double isotropy formalizes the notion of an energy-fair beam sweep, commonly used for initial access in wireless systems.

Aware of the aforementioned issues, focusing on upper-midband \ac{ISAC}, we provide the following contributions:
\begin{itemize}
    \item We introduce the notion of a \textit{doubly isotropic} pilot-combiner. We also provide one possible design procedure for \textit{doubly isotropic} pilot-combiner pairs.
    \item We introduce a \ac{MS} \ac{SAGE} procedure based on the insights of previous work \cite{ch_est_pinto} to tackle the specific problems of \ac{PCE} with hybrid beamformers.
    \item We explore the sensing, communications, and power consumption aspects of different beamforming architectures and converter resolutions.
    \item We analyze the presented results and draw conclusions aimed at orienting the design of \ac{ISAC}-aware radio front end hardware.
\end{itemize}
The paper is organized as follows. In Section~\ref{sec:model}, we introduce the system model to be used throughout the remainder of the work. In Section~\ref{sec:estimation}, we present the considered estimation framework. Then, in Section~\ref{sec:beamforming}, we present an approach for sensing-focused hybrid beamformer design, followed by the introduction of the \ac{MS}-\ac{SAGE} algorithm in Section~\ref{sec:ms_sage}. The sensing and communications performance, as well as the power efficiency aspects of various architectures are compared in Section~\ref{sec:numres}. Finally, the conclusions are drawn in Section~\ref{sec:conclusion}.

\textit{Notation:} We denote vectors and matrices with bold-faced lowercase and uppercase letters, respectively, $\mathbf{x}$ and $\mathbf{X}$. Tensors and operators are denoted by calligraphic uppercase letters, $\mathcal{X}$. Sets (beyond the usual number sets $\mathbb{N}$, $\mathbb{Z}$, $\mathbb{R}$, and $\mathbb{C}$) are denoted either by uppercase calligraphic or Greek letters, e.g., $\mathcal{C}$ or $\Theta$. The indexing of element $ijkl$ of a tensor is denoted by $[\mathcal{X}]_{ijkl}$, similarly defined for other tensor dimensions. Indexing all elements over a particular dimension is denoted by ``$:$'', e.g., $[\mathcal{X}]_{::kl}$ is a matrix defined by the elements over the first and second dimensions of $\mathcal{X}$. The imaginary unit is denoted by $j$ (when $j$ is not used as an index). The Hadamard product is denoted by $\odot$. The Frobenius norm of a tensor is denoted by $\|\cdot\|_F$. The ceiling function (the smallest integer larger than or equal to $x$) is denoted by $\lceil x \rceil$. The indicator function of set $\mathcal{S}$ is denoted by $\mathcal{I}_{\mathcal{S}}(x)$, equal to 1 if $x\in \mathcal{S}$ and 0 otherwise. \textcolor{blue}{Other operators are defined in their usual sense.}

\vspace{-0.4cm}
\section{System Model}\label{sec:model}
We consider a general bistatic sensing scenario. The modeling is divided into three parts. First, we consider the channel model with hybrid beamforming and ideal \ac{Rx} and \ac{Tx}. Then, we include the quantization and amplifier distortion effects. Lastly, we apply the hardware impairments to the ideal model, yielding the full system model.

\vspace{-0.5cm}
\subsection{Ideal Hardware}
Assume that an \ac{OFDM} waveform is transmitted through a double directional time-varying MIMO channel. The received signal is a sum of $L$ narrowband and far-field multipath components \cite{zhang_enabling_jsc}
\begin{equation}
    y_{ntm}  =  \sum^{L}_{\ell=1} y^\ell_{ntm}  + w_{ntm}, \label{eq:rec_sig}
\end{equation}
where $n$, $t$, and $m$ denote the subcarrier, \ac{OFDM} symbol, and receive \ac{RF} chain index, respectively. The contribution of each path is denoted by $y^\ell_{ntm}$. At the \ac{Rx}, the signal is corrupted by \ac{AWGN} of variance $N_0$ represented by $w_{ntm}\sim\mathcal{CN}(0,N_0)$. Denote by $N_\text{C}$ and $N_\text{S}$ the number of subcarriers and OFDM symbols in a pilot frame, respectively. Also consider $D_\text{R}$ receive RF-chains and $N_\text{R}$ receive antennas, and, similarly, $D_\text{T}$ transmit RF-chains and $N_\text{T}$ transmit antennas. \par

Assuming the antennas are arranged in a half-wavelength spaced \ac{ULA} structure, then an element of the channel tensor of multipath $\ell$ is given by 
\begin{equation}
    h^\ell_{ntuv}  =  b_\ell e^{j(n\omega_{1\ell}+t\omega_{2\ell} + u\psi_\ell + v\varsigma_\ell)},
\end{equation}
where $u$ and $v$ are the \ac{Rx} and \ac{Tx} antenna indices, respectively, and $j$ is the imaginary unit. For the multipath $\ell$, $b_{\ell}$ is the complex coefficient of the path and $\omega_{1\ell}$, $\omega_{2\ell}$, $\psi_\ell$, and $\varsigma_\ell$ are angular frequencies related to the \ac{ToF}, Doppler shift, \ac{AoA}, and \ac{AoD}, respectively. These angular frequencies are given by
\begin{multicols}{2}\noindent 
    \begin{align}
        \omega_{1\ell} &= -2\pi(\tau_{\ell}+\tau_\text{o})f_\text{sc},\\
        \omega_{2\ell} &= 2\pi(f^\text{D}_\ell + f_\text{o}) T_\text{s},\!
    \end{align}
    \begin{align}
        \psi_{\ell} &= -\pi\sin(\phi_\ell), \\
        \varsigma_{\ell} &= -\pi\sin(\theta_\ell).
    \end{align}
\end{multicols}\noindent 
The propagation delay and the Doppler frequency of the path are denoted by $\tau_{\ell}$ and $f^D_\ell $, respectively. The clock timing and carrier frequency offsets between the \ac{UE} and the \ac{BS} are respectively given by $\tau_{o}$ and $f_{o}$. The \ac{OFDM} symbol length is represented by $T_s$. Also, the angles of arrival and departure are given by $\phi_\ell$ and $\theta_\ell$, respectively.\par

\begin{figure*}[!tb]
    \includegraphics[width=\textwidth]{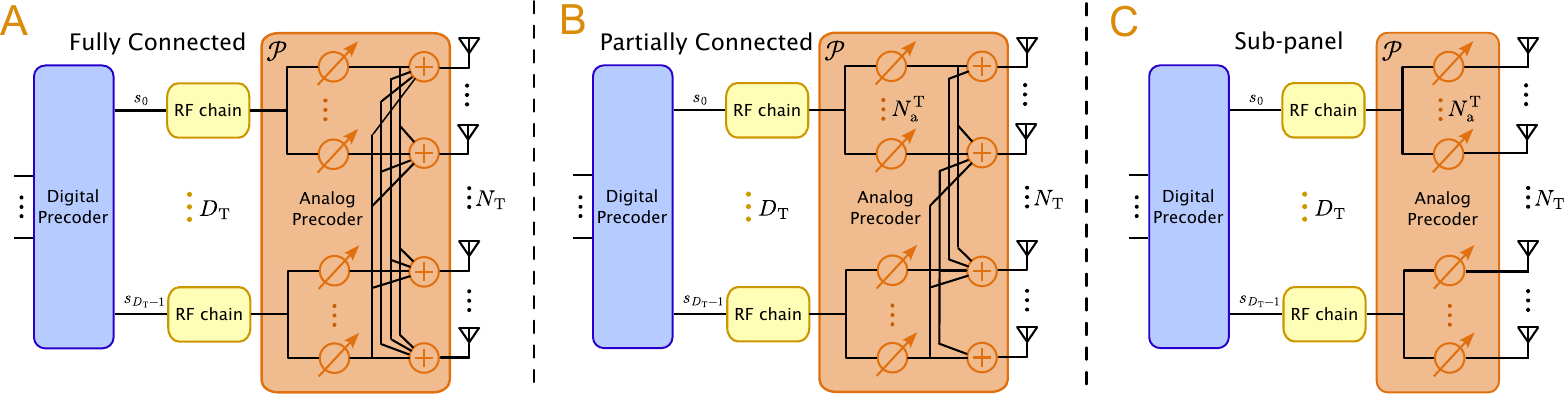}
    \caption{Graphical representation of the fully connected (A), partially connected (B), and sub-panel based (C) hybrid beamformer transmitter topologies. The equivalent receiver topologies are defined in reciprocal fashion. The term \textit{sub-panel} is used as an analogy for the 2D case.}
    \label{fig:hyb_bf}
\end{figure*}

Consider a \ac{Rx}-\ac{Tx} pair equipped with hybrid beamformers such as exemplified in Fig.~\ref{fig:hyb_bf}, then the received signal tensor of multipath $\ell$ is given by
\begin{multline} \label{eq:rx_sig}
    y^\ell_{ntm} = \sum^{N_\text{R}-1}_{u=0} \sum^{N_\text{T}-1}_{v=0} r_{mtu} x_{ntv} h^\ell_{ntuv} \\
     = b_\ell r_{mt}(\psi_\ell) x_{nt}(\varsigma_\ell) e^{j(n \omega_{1\ell} + t\omega_{2\ell})},
\end{multline} 
in which 
\begin{multicols}{2}\noindent 
    \begin{equation}
        r_{mt}(\psi) = \sum^{N_\text{R}-1}_{u=0} r_{mtu} e^{ju\psi}, 
    \end{equation}
    \begin{equation} \label{eq:x_sigma}
        x_{nt}(\varsigma) = \sum^{N_\text{T}-1}_{v=0} x_{ntv} e^{jv\varsigma},  
    \end{equation}
\end{multicols}\noindent 
where $r_{mtu}$ represents the coefficients of the combiner at \ac{OFDM} symbol time $t$ related to \ac{RF} chain $m$ and receive antenna $u$, and $x_{ntv}$ represents the transmitted pilot at subcarrier~$n$, symbol $t$, and transmit antenna $v$. \par

The components $y_{ntu}$, $r_{mtu}$, $x_{ntv}$, and the channel~$h_{ntuv}$ are arranged as complex-valued tensors. Thus, $y_{ntu}$ constitutes the elements of the received signal tensor \mbox{$\mathcal{Y}\in\mathbb{C}^{N_\text{C} \times N_\text{S} \times D_\text{R}}$} and $y^\ell_{ntm}$ are the components of $\mathcal{Y}_\ell$. Likewise, $\mathcal{R}\in\mathbb{C}^{D_\text{R}\times N_\text{S} \times N_\text{R}}$ denotes the combiner tensor, \mbox{$\mathcal{X}\in\mathbb{C}^{N_\text{C}\times N_\text{S} \times N_\text{T}}$} the pilot signal tensor, and \mbox{$\mathcal{H}\in\mathbb{C}^{N_\text{C} \times N_\text{S} \times N_\text{R} \times N_\text{T}}$} the channel tensor, which is comprised of the sum of $L$ multipath channel tensors $\mathcal{H}_\ell$ with components $h^\ell_{ntuv}$. Explicitly,
\begin{equation} \label{eq:channel_tensor}
    h_{ntuv} = [\mathcal{H}]_{ntuv} = \sum^L_{\ell=1} [\mathcal{H}_\ell]_{ntuv} = \sum^L_{\ell=1} h^\ell_{ntuv}.
\end{equation} \par

As an example, assuming $N_\text{R} = 6$ and $D_\text{R}=2$, then a combiner $[\mathcal{R}]_{::t}$ is a $\mathbb{C}^{D_\text{R}\times N_\text{R}}$ matrix of the form
\begin{equation}
    [\mathcal{R}]_{::t} =
    \begin{bmatrix}
        r_{00t} & r_{01t} & r_{02t} & r_{03t} & r_{04t} & r_{05t} \\
        r_{10t} & r_{11t} & r_{12t} & r_{13t} & r_{14t} & r_{15t}
    \end{bmatrix}. \label{eq:fully_connected_combiner}
\end{equation}
Assume similarly that $x_{ntv}$ consists of a pilot sequence with $D_\text{T}$ streams which are transmitted over $N_\text{T}$ transmit antennas through a hybrid beamforming scheme. Consider, for example, that $D_\text{T}=3$ and $N_\text{T}=4$, then we have
\begin{equation}
    [\mathcal{X}]_{nt:} = [\mathcal{P}]_{::t} [\mathcal{S}]_{nt:}
    = \begin{bmatrix}
        p_{00t} & p_{01t} & p_{02t} \\
        p_{10t} & p_{11t} & p_{12t} \\
        p_{20t} & p_{21t} & p_{22t} \\
        p_{30t} & p_{31t} & p_{32t} 
    \end{bmatrix}
    \begin{bmatrix}
        s_{nt0} \\ s_{nt1} \\ s_{nt2}
    \end{bmatrix},
\end{equation}
where $\mathcal{P}$ is the $\mathbb{C}^{N_\text{T} \times D_\text{T}\times N_\text{S}}$ analog precoding tensor and \mbox{$\mathcal{S}\in\mathbb{C}^{N_\text{C}\times N_\text{S} \times D_\text{T}}$} denotes the tensor of pilot symbols. In general, the transmitted signal is given by
\begin{equation} \label{eq:x_from_streams}
    x_{ntv} = \sum^{D_\text{T}-1}_{d=0} p_{vdt} s_{ntd},
\end{equation}
where $p_{vdt}$ are the elements of the precoding tensor $\mathcal{P}$ and $s_{ntd}$ are the elements of the pilot symbols tensor $\mathcal{S}$.
\par

The transmit and receive arrays are subdivided into subarrays of size $N^\text{R}_a$ and $N^\text{T}_a$, respectively. The subarrays are not necessarily disjoint, that is, an antenna may be part of two subarrays at the same time such as in configurations A and B of Fig.~\ref{fig:hyb_bf}. As in the previous example in (\ref{eq:fully_connected_combiner}), if all the elements of $[\mathcal{R}]_{::t}$ are nonzero, then it describes a \textit{fully connected} combiner as shown in configuration A of Fig.~\ref{fig:hyb_bf}. As in configuration B of Fig.~\ref{fig:hyb_bf}, a \textit{partially connected} configuration with $N^\text{R}_a = 4$ has the following structure
\begin{equation}
    [\mathcal{R}]^\text{PC}_{::t} = 
    \begin{bmatrix}
        r_{00t} & r_{01t} & r_{02t} & r_{03t} & 0 & 0 \\
        0 & 0 & r_{12t} & r_{13t} & r_{14t} & r_{15t}
    \end{bmatrix}.
\end{equation}
A \textit{sub-panel} topology, as in configuration C of Fig.~\ref{fig:hyb_bf}, with $N^\text{R}_a = 3$, has a combining matrix given by
\begin{equation}
    [\mathcal{R}]^\text{SP}_{::t} =
    \begin{bmatrix}
        r_{00t} & r_{01t} & r_{02t} & 0 & 0 & 0 \\
        0 & 0 & 0 & r_{10t} & r_{11t} & r_{12t}
    \end{bmatrix}. \label{eq:partially_connected_combiner}
\end{equation}
The structures for the defined precoder topologies are defined analogously.

\vspace{-0.4cm}
\subsection{Hardware Impairment Model}
We will first describe the modeling of the amplifier nonlinearities, which will be used for the \ac{LNA} and \ac{PA}. Then, we detail the quantization model, which shall cover the modeling for both the \ac{ADC} and \ac{DAC}. The phase noise model is then introduced. Finally, the hardware impairment models are merged in the full system model. We consider the wideband effects of \ac{IQ} imbalance and filter ripple to be negligible.

\subsubsection{Amplifier Nonlinearity}
Amplifiers typically display a wide variety of nonideal behaviors, such as nonlinear distortion, saturation, and memory effects. For simplicity, we choose to model both the \ac{PA} and \ac{LNA} as memoryless systems. The functions representing the memoryless nonlinearities of the \ac{PA} and \ac{LNA}, denoted by $\mathcal{A}^\text{PA}$ and $\mathcal{A}^\text{LNA}$, respectively, are applied in elementwise fashion to a vector of time domain samples. \par
% However, it is convenient to represent their effect in the subcarrier domain. Thus, we respectively denote the outputs of the \ac{PA} and \ac{LNA} as $\widecheck{s}_{ntd} = \mathcal{A}^\text{PA}_n(\mathcal{S}_{:td})$ and $\widecheck{y}_{ntm} = \mathcal{A}^\text{LNA}_n(\mathcal{Y}_{:tm})$. The $\mathbb{C}^{N_\text{C}}\rightarrow\mathbb{C}$ functions $\mathcal{A}^\text{PA}_n$ and $\mathcal{A}^\text{LNA}_n$ are defined as
% \begin{align}
%     \mathcal{A}^\text{PA}_n(\mathcal{S}_{:td}) &= [\mathbf{F}^H \mathcal{A}^\text{PA}(\mathbf{F} [\mathcal{S}]_{:td})]_n, \label{eq:pa_nonlin} \\
%     \mathcal{A}^\text{LNA}_n(\mathcal{Y}_{:tm}) &= [\mathbf{F}^H \mathcal{A}^\text{LNA}(\mathbf{F} [\mathcal{Y}]_{:tm})]_n, \label{eq:lna_nonlin}
% \end{align}
% where $\mathbf{F}$ is the \ac{DFT} matrix. One can see that, in (\ref{eq:pa_nonlin}), the transmit streams are first represented in the time domain, then subjected to the \ac{PA} nonlinearity, and finally converted back to subcarrier domain. The same process happens in (\ref{eq:lna_nonlin}).\par

It is important to state that, since \ac{DPD} is not being explicitly modeled in this work, the \ac{PA} nonlinearity actually represents the residual nonlinearity after \ac{DPD}. Thus, we state the following assumption:

\begin{assumption}[Amplifier Residual Nonlinearities]\label{assumption:res_nonlin}
    The \ac{LNA} nonlinearity and the \ac{PA} residual nonlinearity after \ac{DPD} are well modeled by a sigmoidal function. 
\end{assumption}
Assumption~\ref{assumption:res_nonlin} is based on the usual results of standard \ac{DPD} algorithms applied to amplifiers modeled as memoryless nonlinearities (or with relatively weak memory effect). In those scenarios, the small amplitude response of the compensated amplifier is approximately linear. The effectiveness of \ac{DPD} progressively deteriorates as the amplifier approaches saturation. Thus, we claim that Assumption~\ref{assumption:res_nonlin} is a valid simplification for many contexts. Nevertheless, considering a more realistic model for the \ac{PA} with \ac{DPD} is a necessary next step which is left for future work.

As for the specific model used for $\mathcal{A}^\text{PA}$ and $\mathcal{A}^\text{LNA}$, we use a scaled arctangent function. Given a \ac{Tx} backoff factor $\nu_T$, and a \ac{PA} nonlinearity parameter $\kappa_\text{PA}$, we model the \ac{PA} as
\begin{equation}
    \mathcal{A}^\text{PA}(x) = \nu_T \kappa_\text{PA} \! \left( \! \arctan\left( \frac{\Re{x}}{\nu_T \kappa_\text{PA}} \right) \! + \! j\arctan\left( \frac{\Im{x}}{\nu_T \kappa_\text{PA}} \right) \! \right).
\end{equation}
The function is perfectly linear in the neighborhood of zero amplitude, and progressively compresses the signal as the amplitude increases. A similar model is used for the \ac{LNA}, but with parameters $\nu_R$ and $\kappa_\text{LNA}$.

\subsubsection{Quantization} \label{subsec:quantizers}
As in \cite{adc_pinto}, each quantizer is fully described by its quantization regions and representation levels. The quantization regions are denoted by the tuple of sets \mbox{$\Theta = (\Theta_1,\dots,\Theta_K)$}, which is also a partition of the real line, i.e., $\bigcup^K_{k=1} \Theta_k = \mathbb{R}$, $\Theta_k \neq \emptyset$, and $\Theta_{k_1}\cap\Theta_{k_2} = \emptyset$ when ${k_1} \neq {k_2}$. We further require that $\Theta_k$ be connected, i.e., each set cannot be represented as the union of two or more disjoint non-empty open subsets, and that $\sup \Theta_{k_1}\leq\inf\Theta_{k_2}$ if $k_1<k_2$, with equality only when $k_2 = k_1+1$. The representation levels are denoted by the set $\mathcal{C}=\{c_k\}^K_{k=1}$, which has an isomorphism \mbox{$q(\Theta_k) = c_k$} between it and the quantization regions, such that $c_k \in q^{-1}(c_k)$. Therefore, a quantizer $\mathcal{Q}$ is fully defined by $(\Theta,\mathcal{C})$, with the quantization operation being \vspace{-0.25cm}
\begin{equation} \label{eq:quantizer}
    %\mathcal{Q}(x) = \sum_{z \in \Theta} \mathcal{I}_{z}(x) q(z) = \sum^M_{m=1} \mathcal{I}_{\Theta_m}(x)c_m,
    \mathcal{Q}(x) = \sum^K_{k=1} \mathcal{I}_{\Theta_k}(\Re{x})c_k + j\sum^K_{k=1} \mathcal{I}_{\Theta_k}(\Im{x})c_k.
\end{equation}

We consider uniform and optimal $K$-level quantizers. Uniform quantizers are parameterized uniquely by their ends-of-scale $(\Theta_\text{min},\Theta_\text{max})$ and number of bins $K$, such that
\begin{equation}
    \Theta_k = (\Theta_\text{min} + (k-2)\Delta, \Theta_\text{min} + (k-1)\Delta],
\end{equation}
for $k=2,\dots,K-1$, with the edge cases $\Theta_1 = (-\infty,\Theta_\text{min}]$ and $\Theta_K = (\Theta_\text{max},\infty)$, and $\Delta = \frac{\Theta_\text{max} - \Theta_\text{min}}{K-2}$. We assume that $\Theta_\text{max} = -\Theta_\text{min}>0$. Letting the received signal have zero mean and known variance, define $\Theta_\text{max} = \nu\sigma_\text{in}$, where $\nu\in\mathbb{R}^*_+$ is the backoff factor, and $\sigma_\text{in}$ is the standard deviation of the input signal of the quantizer (assuming that the input is at least a second order stationary random process). Assuming the in-phase and quadrature branches have the same distribution, an optimal quantizer is computed by applying the \ac{LM} algorithm to a dataset with the real and imaginary values from multiple input realizations.

The quantization operation works in symmetrical fashion for the \ac{ADC} and \ac{DAC}. This means that both $\mathcal{Q}^\text{DAC}$ and $\mathcal{Q}^\text{ADC}$ describe operationally identical (for similar quantization regions and representation points), but conceptually different functions. Namely, the \ac{DAC} maps a digital sample to an analog time domain sample and the \ac{ADC} maps an analog time domain sample to a digital sample. The transition from analog to digital domain (and vice versa) is performed implicitly (i.e., we do not model analog signals as continuous functions) \textcolor{blue}{assuming reconstruction at $f_\text{smp} = N_\text{C} f_\text{sc}$ with an ideal anti-alias low-pass filter and with Nyquist sampling at \ac{Rx}. Because the continuous domain is never modeled, this also means that spectral regrowth caused by nonidealities gets folded back in-band; thus, representing a pessimistic scenario.}

\subsubsection{Phase Noise}
In the considered carrier frequency range, phase noise is also known to noticeably impact the performance of communications systems. Phase noise is usually characterized by its \ac{PSD} (given in dBc/Hz) which results from the compounding of multiple noise sources \cite{ericsson_phase_noise}, such as the oscillator noise, thermal phase noise, in-band flicker noise, etc. We considered a model adapted from the 29.55~GHz phase noise model in \cite[Section~6.1.10]{3gpp.tr.38.803}. A desired phase noise \ac{PSD} is obtained by converting the aforementioned model from 29.55~GHz to 7.125~GHz by adding $20 \log_{10}(7.125/29.55)$ to the initial \ac{PSD} curve. This conversion is justified by the claim in \cite[Section~6.1.9.5]{3gpp.tr.38.803}, which states that phase noise can increase by 6~dB for each doubling of the carrier frequency.\par

Phase noise is described in time domain by the expression \vspace{-0.15cm}
\begin{equation} \label{eq:phase_noise}
    \textcolor{blue}{\mathbf{x}_\text{PN} = \mathcal{M}(\mathbf{x}) = \mathbf{x} \odot \begin{bmatrix} e^{j\phi_1} & \cdots & e^{j\phi_N} \end{bmatrix}^T,}
\end{equation}
where \textcolor{blue}{$\mathcal{M}$ is the phase noise operator}, $\mathbf{x}_\text{PN}$ is the resulting vector affected by phase noise, $\mathbf{x} \in \mathbb{C}^{N}$ is a generic input of contiguous time domain samples, and $\phi_k$ is the phase noise at index $k$, modeled by filtering a white Gaussian noise sequence with the filter $\mathcal{F}$. In this work, $\mathcal{F}$ is designed as an \ac{IIR} filter parametrized by a specified point $(f_x, p_x)$ in the phase noise \ac{PSD}. This is based on the implementation of the \texttt{PhaseNoise} function in Matlab Communications Toolbox. The numerator of the filter response is given by $a_\text{num} = \sqrt{2 \pi f_x 10^{p_x/10}}$, while the denominator coefficients of the filter are determined recursively $\gamma_i = (i-2.5)\frac{\gamma_{i-1}}{i-1}$,
with $\gamma_1 = 1$, where $i=1,2,\dots,N_\text{coef}$ such that \mbox{$N_\text{coef} = 2^7$}. The result is a random process with a \ac{PSD} which is close to $p_x$ at a frequency offset of $f_x$ from the carrier $f_c$ and decreases at 10dB/decade. The chosen reference point was \mbox{$(p_x, f_x) = (-65 + 20 \log_{10}(7.125/29.55), 1000)$}, in which the value of $-65$dB was approximated visually from \cite[Fig. 6.1.10-2]{3gpp.tr.38.803}. We assume that both \ac{Tx} and \ac{Rx} have similar phase noise behavior.

\vspace{-0.35cm}
\subsection{Full System Model}
We will now combine the nonideal hardware and ideal signal models to obtain the complete system model which will be used throughout the remainder of this work. By combining (\ref{eq:x_from_streams}), (\ref{eq:quantizer}), (\ref{eq:phase_noise}), and the amplifier distortion, we get
\begin{gather} \label{eq:x_nonideal}
    \bar{x}_{ntv} = \sum^{D_\text{T}-1}_{d=0} p_{vdt} \widecheck{s}_{ntd}, \\
    \widecheck{s}_{ntd} = \left[\mathbf{F}^H \mathcal{A}^\text{PA}(\mathcal{M}(\mathcal{Q}^\text{DAC}(\mathbf{F}[\mathcal{S}]_{:td}))) \right]_{n}, \label{eq:s_nonideal}
\end{gather} 
where $\mathbf{F}$ is the \ac{DFT} matrix. Both the quantizer and amplifier functions are applied in an elementwise fashion. The phase noise is applied to the whole vector, due to the filtering operation. One can see that, in (\ref{eq:s_nonideal}), the transmit streams are first represented in the time domain, subjected to the \ac{DAC}, affected by upconversion phase noise, distorted by the \ac{PA} nonlinearities, and converted back to subcarrier domain. The nonideal streams are then precoded in (\ref{eq:x_nonideal}). After that, the pilot is transmitted through the channel. At the receiver, the received signal is amplified by the \acp{LNA}, affected by the downconversion phase noise, and quantized by the \acp{ADC}, yielding
\begin{equation} \label{eq:impaired_rec_sig}
    \bar{y}_{ntm} =  \left[\mathbf{F}^H \mathcal{Q}^\text{ADC}(\mathcal{M}(\mathcal{A}^\text{LNA}(\mathbf{F} [\widecheck{\mathcal{Y}}]_{:tm}))) \right]_n,
\end{equation}
where $\widecheck{\mathcal{Y}}$ is the $\mathbb{C}^{N_\text{C} \times N_\text{S} \times D_\text{R}}$ tensor with elements given by
\begin{equation} 
    \widecheck{y}_{ntm} = \sum_{u,v} r_{mtu} \bar{x}_{ntv} h_{ntuv} + w_{ntm}.
\end{equation}

\vspace{-0.6cm}
\subsection{Power Consumption Model}
Finally, we present the considered model for the total power consumption of each \ac{RF} chain. This model will be used to evaluate the tradeoff between sensing plus communications performance and power consumption. The expressions for the power consumption of individual components are drawn from \cite{power_expressions}. The considered \ac{ADC} consumption model is given by
\begin{equation}
    P_\text{ADC}(N^\text{bits}_\text{ADC}) = c_\text{ADC} 2^{N^\text{bits}_\text{ADC}},
\end{equation}
where $c_\text{ADC} = 10^{-3}$ is the \ac{ADC} power coefficient and $N^\text{bits}_\text{ADC}$ is the resolution of the \ac{ADC}.  The \ac{ADC} power coefficient was based on the value extracted from \cite{orig_adc_power_model}, where is assumed to be $c_\text{ADC} = 10^{-4}$ for a signal bandwidth of 1~MHz. In this work we increase this value by an order of magnitude as an attempt to better represent the possible power consumption model of systems with larger bandwidth. Regarding the \ac{DAC}, its power consumption is given by
\begin{equation}
    P_\text{DAC}(N^\text{bits}_\text{DAC}) = c^a_{\text{DAC}} 2^{N^\text{bits}_\text{DAC}} + c^b_{\text{DAC}} N^\text{bits}_\text{DAC} f_\text{smp},
\end{equation}
where $c^a_{\text{DAC}} = 1.5 \cdot 10^{-5}$ and $c^b_{\text{DAC}} = 9\cdot 10^{-12}$ are the \ac{DAC} power coefficients, $N^\text{bits}_\text{DAC}$ is the resolution of the \ac{DAC}, and $f_\text{smp}$ is the sampling frequency. 

The \ac{PA} power consumption model is essentially composed of the power amplifier bias and the transmit signal power divided by the amplifier efficiency, which is also a function of the transmit power. The \ac{PA} consumption is given by
\begin{equation} 
    P_\text{PA} = P_\text{bias} + P_\text{chain}/\textcolor{blue}{\beta_\text{PA}},
\end{equation}
where $P_\text{bias}$ is the bias power consumption of the amplifier, and \textcolor{blue}{$\beta_\text{PA}(P_\text{chain})$} is the efficiency of the amplifier. We consider that all the \acp{PA} operate at $\textcolor{blue}{\beta_\text{PA}}=0.3$ efficiency with a bias consumption of $P_\text{bias}=1$~W. The total \ac{RF} chain power consumption is a function of the number of \ac{RF} chains, as well as the \ac{DAC} and \ac{ADC} resolutions
\begin{equation} \label{eq:pwr_model}
    P_\text{tot}(D,N^\text{bits}_\text{DAC},N^\text{bits}_\text{ADC}) = D\left( P_\text{PA} + P_\text{DAC} +  P_\text{ADC} \right). 
\end{equation}

\vspace{-0.35cm}
\section{Estimation Framework} \label{sec:estimation}
In this section, we will detail the \ac{PCE} approach that will be used throughout the remainder of the paper. We begin by describing the assumptions on the noise components, then we proceed by presenting the description and relevant theory of the \ac{SAGE} algorithm.

\begin{assumption}[AWGN Impairment Noise in Estimation] \label{assumption:noise_modeling}
    Given (\ref{eq:rec_sig}) and (\ref{eq:impaired_rec_sig}), the impaired received signal is written as \vspace{-0.2cm}
    \begin{equation}
        \bar{y}_{ntm} = y_{ntm} + z_{ntm},
    \end{equation}
    where $z_{ntm}$ is a noise component representing the effects of the \textcolor{blue}{nonidealities} in the received signal as well as thermal noise, with its respective tensor being $\mathcal{Z}$. In the estimation process, for mathematical tractability, we assume that the elements of $\mathcal{Z}$ are \ac{i.i.d.} Gaussian with variance $N^z_0$. We assume also that $\mathcal{Z}$ is independent from $\mathcal{Y}$.
\end{assumption}
Clearly, Assumption~\ref{assumption:noise_modeling} ignores the statistical intricacies of the considered hardware impairments. However, obtaining explicit expressions for the resulting impairment error distributions is extremely challenging or even mathematically intractable, except in very simplified scenarios. Thus, at least for an initial treatment of the problem, Assumption~\ref{assumption:noise_modeling} is a necessary simplification. It can be observed from numerical studies that (for sufficiently large $N_\text{C}$ and sufficiently dispersive channels \cite{adc_pinto}) the elements of $\mathcal{Z}$ are approximately uncorrelated and have zero mean, unimodal marginals.

In general, differently from the \ac{BD} approach, $\mathcal{Z}$ is colored and depends on $\mathcal{Y}$. However, determining the correlation matrix of $\mathcal{Z}$ is untractable for practical estimation, since it depends on the hardware impairments and on the channel itself. Assuming an ideal transmitter, we have shown in \cite{adc_pinto} that the receiver quantization error can be accurately approximated as Gaussian when the channel is dispersive, i.e., the maximum power contribution of any distinct path is proportionally negligible to the sum of all paths. It was also shown in \cite[Fig.~2]{adc_pinto} that, using optimal \mbox{1-bit} quantizers in a \ac{LOS} channel\footnote{\Ac{LOS} dominant channels are a worst case scenario, since adding independent paths makes the signal closer to satisfying the central limit theorem.}, the receiver quantization error can be already considered approximately Gaussian. Although Assumption \ref{assumption:noise_modeling} may seem strong, it will be numerically observed in Section~\ref{sec:numres} that it nevertheless leads to good estimation performance. We note that Assumption \ref{assumption:noise_modeling} is only used in the context of the derivation of the estimator; the system model used to generate numerical results still follows the description in Section \ref{sec:model}.

%\textcolor{red}{Check Emil's paper on SE and quantization mMIMO to add worst case scenario when noise is white}. 
% This approach can be contrasted with the Bussgang decomposition \cite{bussgang}, which allows us to model the $\bar{y}_{ntm}$ as
% \begin{equation}
%     \bar{y}_{ntm} = \sum_{nm} g_{nm} y_{ntm} + \widecheck{z}_{ntm}
% \end{equation}
% where $g_{nm}$ is the kernel of an uncorrelating transform that makes $\widecheck{z}_{ntm}$ uncorrelated to both $y_{ntm}$ and $\bar{y}_{ntm}$. Note that the transform is performed over the subcarriers and \ac{Rx} \ac{RF} chains, this is done because the signals are already independent over $t$ by definition.

\vspace{-0.35cm}
\subsection{Parametric Estimation}
We perform \ac{PCE} based on the \ac{SAGE} procedure, a maximum likelihood method. Let $\boldsymbol{\xi} = (\boldsymbol{\xi}_1,\dots,\boldsymbol{\xi}_L)$ denote the tuple including all estimation parameters. Then, omitting constant terms, the resulting log-likelihood function is given by \cite{kay_mle}
\begin{equation}  
     \log \mathcal{L}(\boldsymbol{\xi}; \bar{\mathcal{Y}}) = -\frac{\sum_{ntm} \left| \bar{y}_{ntm} - y_{ntm}(\boldsymbol{\xi}) \right|^2}{N^z_0} + \dots, %= \frac{\|\mathcal{Y} - \mathcal{Y}(\boldsymbol{\xi}) \|^2_F}{N_0}  + \dots
\end{equation}
where $y_{ntm}(\boldsymbol{\xi})$ is computed from (\ref{eq:rec_sig}). The summation indices have been omitted and, henceforth, summations and products over $m/n/t/v$ go from 0 to $N_{\text{R}/\text{C}/\text{S}/\text{T}}-1$, unless otherwise indicated. Given $L$, \ac{ML} estimation is equivalent to solving \vspace{-0.15cm}
\begin{equation}\label{eq:opt_problem} 
    \min_{\boldsymbol{\xi}} f(\boldsymbol{\xi}), \text{ where } f(\boldsymbol{\xi}) =  \! \sum_{ntm} \left| \bar{y}_{ntm} - \! \sum^L_{\ell=1} y^\ell_{ntm}(\boldsymbol{\xi}_\ell) \right|^2.
\end{equation} 
Equivalently, $f(\boldsymbol{\xi}) = \left\| \bar{\mathcal{Y}} - \! \sum^L_{\ell=1} \mathcal{Y}_\ell(\boldsymbol{\xi}_\ell) \right\|^2_F$.

The method estimates the parameters of a single path at a time, denoted by the index $\ell'$. This leads to an optimization subproblem given by \cite{ch_est_pinto,sage}
\begin{gather} \label{eq:sage_l_obj}
    g(\boldsymbol{\xi}_{\ell'}) = \frac{ \left| \sum_{ntm} \alpha^{\ell'}_{ntm}(\boldsymbol{\xi}_{\ell'}) \Tilde{y}^{\ell'*}_{ntm} \right|^2 }{\sum_{ntm} \left| \alpha_{ntm}(\boldsymbol{\xi}_{\ell'}) \right|^2 },\\
    \Tilde{y}^{\ell'}_{ntm} = \bar{y}_{ntm} - \sum_{\ell \neq \ell'} b_\ell r_{mt}(\psi_\ell) x_{nt}(\varsigma_\ell) e^{j(n \omega_{1\ell} + t\omega_{2\ell})}, \label{eq:y_tilde} \\
    \alpha^{\ell'}_{ntm}(\boldsymbol{\xi}_{\ell'}) = r_{mt}(\psi_{\ell'}) x_{nt}(\varsigma_{\ell'}) e^{j(n \omega_{1{\ell'}} + t\omega_{2{\ell'}})}.
\end{gather}
If there is accurate knowledge of the \ac{Tx} impairments, $\bar{x}_{ntv}$ can be used instead of $x_{ntv}$ to compute $x_{nt}(\varsigma_\ell)$ in (\ref{eq:y_tilde}) by substituting (\ref{eq:x_nonideal}) in (\ref{eq:x_sigma}).\par

If there is no previous knowledge of the channel tensor, it is reasonable to deliver the same energy to all angles of departure and combine with the same gain towards all angles of arrival. This is an informal statement of the condition 
\begin{gather}
     \frac{\partial}{\partial \psi} \alpha(\psi,\varsigma) = \frac{\partial}{\partial \varsigma} \alpha(\psi,\varsigma) = 0, \label{eq:double_isotropy}\\
    \alpha(\psi,\varsigma) = \sum_{ntm} |r_{mt}(\psi)|^2 |x_{nt}(\varsigma)|^2 
\end{gather}
which describes perfectly energy-fair beam sweeping. The condition in (\ref{eq:double_isotropy}) is satisfied if $\mathcal{R}$ and $\mathcal{X}$ make $\alpha(\psi,\varsigma)$ constant over $\psi$ and $\varsigma$. A combiner-pilot pair $(\mathcal{R},\mathcal{X})$ that satisfies (\ref{eq:double_isotropy}) is called \textit{doubly isotropic}. Note that the double isotropy condition relates exclusively to the energy properties of the whole pilot-combiner frame over the spatial/angular domain. This means that there exist multiple possible doubly isotropic pilot-combiner structures with different properties regarding delay and Doppler identifiability and resolution. Naturally, some approaches exhibit better delay-Doppler ambiguity-function characteristics, but we do not explore this further here. The design of doubly isotropic beamformers will be explored in Section~\ref{sec:beamforming}.\par

\vspace{-0.35cm}
\subsection{Parameter Updates}
The subproblem in (\ref{eq:sage_l_obj}) is solved by updating one parameter at a time, i.e., alternating coordinate descent. Then, after updating every parameter in path $\ell'$, the path coefficient is computed from a closed form expression. The updates for $\omega_{1}$ and $\omega_2$ follow the same structure as detailed in \cite{ch_est_pinto}, with the needed modifications for $r_{mt}$ and $x_{nt}$. While similar to \cite[Eq. (16)]{ch_est_pinto}, the expression for the optimal path coefficient now includes the combiner terms in the denominator %
\begin{equation} \label{eq:b_opt_hybrid} 
    b_{\ell'}(\boldsymbol{\xi}_{\ell'}) = \left(\sum_{ntu} \alpha^{\ell'*}_{ntu} \Tilde{y}^{\ell'}_{ntu}\right) \bigg/ \sum_{ntm} |r_{mt}(\psi_\ell)|^2 |x_{nt}(\varsigma_\ell)|^2.
\end{equation}
Likewise, due to the hybrid combining setup, the expression for the $\psi$ update needs to be modified more extensively. Some details about the $\varsigma$ updates also need to be addressed. We present this information in the following subsections.

\subsubsection{Angle of Arrival Update}
Define the equivalent objective by eliminating the denominator term \vspace{-0.2cm}
\begin{equation} \label{eq:subproblem_no_denom}
    h(\boldsymbol{\xi}_{\ell'}) =  \left| \sum_{ntm} \alpha_{ntm}(\boldsymbol{\xi}_{\ell'}) \Tilde{y}^*_{ntm} \right|^2,
\end{equation}
we will then compute its Fourier coefficients by first finding the Fourier expansion of $\sum_{ntm} \alpha_{ntm}(\boldsymbol{\xi}_{\ell'}) \Tilde{y}^*_{ntm}$ and applying the properties of the Fourier series. Expanding, we get
\begin{multline}
    \sum_{ntm} \!\! \alpha_{ntm}(\boldsymbol{\xi}_{\ell'}) \Tilde{y}^*_{ntm} \!\! = \!\!\! \sum_{ntm} \!\! r_{mt}(\psi_{\ell'}) x_{nt}(\varsigma_{\ell'}) e^{j(n \omega_{1{\ell'}} + t\omega_{2{\ell'}})}\Tilde{y}^*_{ntm} \\
    = \sum_{ntm} \sum^{N_\text{R}-1}_{u=0} r_{mtu} e^{ju\psi'} x_{nt}(\varsigma_{\ell'}) e^{j(n \omega_{1{\ell'}} + t\omega_{2{\ell'}})}\Tilde{y}^*_{ntm}.
\end{multline}
Defining $\gamma_{u} = \sum_{ntm} r_{mtu} x_{nt}(\varsigma_{\ell'}) e^{j(n \omega_{1{\ell'}} + t\omega_{2{\ell'}})}\Tilde{y}^*_{ntm}$ we get 
\begin{align}
    \sum_{ntm} \alpha_{ntm}(\boldsymbol{\xi}_{\ell'}) \Tilde{y}^*_{ntm} &= \sum^{N_\text{R}-1}_{u=0} \gamma_{u} e^{ju\psi'}.
\end{align}
The rest follows as usual from the Fourier properties of multiplication by the complex conjugate as in \cite[Sec.~IV-A]{ch_est_pinto}. The optimal parameter is solved through companion matrix methods as described in \cite[Sec.~IV-D]{ch_est_pinto}

\subsubsection{Angle of Departure Update} \label{subsubsec:sigma_update}
The companion matrix root-finding approach from \cite[Sec.~IV-D]{ch_est_pinto} relies on the optimization subproblem being a periodogram, i.e., having a Fourier series structure. Assuming a doubly isotropic combiner-pilot pair, we have two possible approaches for updating $\varsigma$. The used approach depends on whether or not the \ac{Tx} hardware impairments are known:
\begin{enumerate}
    \item If the impairments are not known, we can assume the transmitted pilot to be $\mathcal{X}$, which is isotropic, leading to expressions like those in \cite[Sec.~IV-C]{ch_est_pinto} and allowing for exact line-search using companion matrix root-finding. \label{item:not_known_impairments}
    \item If the impairments are known, then the transmitted pilot is $\bar{\mathcal{X}}$, which is generally no longer isotropic. In that case, the denominator of (\ref{eq:sage_l_obj}) is no longer constant in $\varsigma$ and numerical line search must be performed. \label{item:known_impairments}
\end{enumerate}
If we do not have a good enough model for the transmit impairments, then, as described in item ``\ref{item:not_known_impairments})'', we can assume the transmitted pilot to be the isotropic $\mathcal{X}$. This, however, comes at the cost of increased impairment noise, since the \ac{Tx} impairments should now be treated as noise. A major problem of impairment-caused noise is that it is a function of the transmitted and received signal, thus, differently from thermal noise, its power is mostly proportional to the transmit and received power respectively. If the transmit impairments are well known, as in item ``\ref{item:known_impairments})'', then we forego the isotropy assumption, but gain the advantage of reduced impairment noise. This is generally the better approach in terms of estimation accuracy (whenever possible). %Both approaches will be studied numerically in Section \ref{sec:numres}.

\section{Doubly Isotropic Beamformer Design} \label{sec:beamforming}

In this section, we propose methods of generating doubly isotropic combiner-pilot pairs, i.e., pilot-combiner pairs that lead $\alpha(\psi,\varsigma)$ to be invariant with respect to $\psi$ and $\varsigma$. Initially, we borrow from traditional beam sweeping strategies in which the receive beam only changes once the \ac{Tx} went through all transmit beams in the codebook, thus all \ac{Rx}-\ac{Tx} beam pairs are explored. \par

\vspace{-0.35cm}
\subsection{Valid Subarray Structures}
Before going into the design itself, we must clarify our approach towards defining subarray structures. Consider the example case with $N_\text{R}=8$ and $D_R=2$, and assume disjoint subarrays of size $N^R_a=4$. In the described case, the combiner $\mathcal{R}$ is given by
\begin{equation}
    [\mathcal{R}]_{::t} = 
    \begin{bsmallmatrix}
        a^\text{R}_{00t} & a^\text{R}_{01t} & a^\text{R}_{02t} & a^\text{R}_{03t} &  0 & 0 & 0 & 0 \\
        0 & 0 & 0 & 0 & a^\text{R}_{10t} & a^\text{R}_{11t} & a^\text{R}_{12t} & a^\text{R}_{13t} 
    \end{bsmallmatrix}.
\end{equation}
As another example, assuming non-zero elements, (\ref{eq:fully_connected_combiner}) corresponds to the fully connected hybrid beamformer. \par

Not all $(D_\text{R}, N^\text{R}_a, N_\text{R})$ tuples lead to valid ``well distributed'' arrays, in which the overlap between adjacent subarrays is always the same. It can be shown that, for well distributed arrays, given $N^\text{R}_a$ and $N_\text{R}$ values, $D_\text{R}$ must belong to the set\footnote{The set $\mathcal{D}_\text{R}$ is not valid for the $N_\text{R}=N^\text{R}_a$ case, in which non-redundant values of $D_\text{R}$ belong to $\{1,\dots,N_\text{R}\}$.} 
\begin{equation}
    \mathcal{D}_\text{R}  =  \left\{d  \in  \mathbb{N}  :  (N_\text{R}-N^\text{R}_a)|(d-1) \text{ and } \frac{N_\text{R}-N_a}{d-1} \! \leq \! N_a \right\},
\end{equation}
where $a|b$ denotes that $b$ is a divisor of $a$. This constraint ensures that the subarrays cover the entire array and that adjacent subarrays have the same overlapping antennas. The same statement is valid by symmetry to the transmit array.\par

\vspace{-0.35cm}
\subsection{Combiner-Pilot Design -- Analog Precoding} \label{subsec:comb_pilot_analog}
Now we can properly go into the combiner-pilot design. Let us first assume that the \ac{Tx} has an analog beamformer and the \ac{Rx} is equipped with a hybrid combiner. It is known from the design of isotropic pilots with analog beamforming that at least $N_\text{T}$ pilot symbols are needed for an isotropic pilot \cite{ch_est_pinto}. This is the case because the beamformer must be matched to at least $N_\text{T}$ equidistant (e.g., $\Bar{\varsigma}_k = \frac{2\pi}{K} k$) spatial frequencies $\{\Bar{\varsigma}_k\}_{k=1,\dots, K}$, with $K\geq N_\text{T}$, to cover all angles equally (i.e., \ac{DFT} codebook). In conventional beam sweeping, we sweep the combiner over multiple angles of arrival in the codebook while the beamformer is kept constant for each $\Bar{\varsigma}_k$.\par

Assume that the domain of $(n,t,m)$ may be partitioned into $K$ subsets $\mathcal{M}_k$ such that
\begin{equation}
    \frac{\partial}{\partial \psi} \!\!\!\!\!\! \!\!\!\!\!\! \! \sum_{\quad\quad(n,t,m)\in \mathcal{M}_k} \!\!\!\!\!\! \!\!\!\!\!\! \! |r_{mt}(\psi)|^2 = 0. \label{eq:aoa_isotropy}
\end{equation}
Also assume that $|x_{nt}(\varsigma)|^2 = |x_k(\varsigma)|^2$ remains constant during the tuples $(n,t,m)$ belonging to each $\mathcal{M}_k$. Then, we can write
\begin{multline}
    \sum_{ntm} |r_{mt}(\psi_\ell)|^2 |x_{nt}(\varsigma_\ell)|^2 =  \sum^K_{k=1} \!\!\!\!\!\! \!\!\!\!\! \sum_{\quad\quad(n,t,m)\in \mathcal{M}_k}\!\!\!\!\!\! \!\!\!\!\!\! \! |r_{mt}(\psi_\ell)|^2 |x_{nt}(\varsigma_\ell)|^2 \\
    = \sum^K_{k=1} |x_k(\varsigma)|^2 \!\!\!\!\!\! \!\!\!\!\!\!\! \sum_{\quad\quad(n,t,m)\in \mathcal{M}_k}\!\!\!\!\!\! \!\!\!\!\!\! \! |r_{mt}(\psi_\ell)|^2.
\end{multline}
Now, if 
\begin{equation}
    \sum_{\quad\quad(n,t,m)\in \mathcal{M}_k}\!\!\!\!\!\! \!\!\!\!\!\! \! |r_{mt}(\psi_\ell)|^2 = r\text{ for all }k=1,\dots,K,
\end{equation}
that is, $r_{mt}(\psi)$ satisfies (\ref{eq:aoa_isotropy}) while also yielding the same value for all $\mathcal{M}_k$ summations, then we get 
\begin{multline}
    \sum_{ntm} |r_{mt}(\psi_\ell)|^2 |x_{nt}(\varsigma_\ell)|^2 = r \sum^K_{k=1} |x_k(\varsigma)|^2 \\
    \Rightarrow  r \frac{\partial}{\partial \varsigma} \sum^K_{k=1} |x_k(\varsigma)|^2 = 0 \text{ if } |x_k(\varsigma)|^2 \text{ is isotropic}.
\end{multline}
For the specifics of designing isotropic combiners, we can borrow from our knowledge in the design of isotropic pilots for analog beamformers. If a receiver has $D_\text{R}$ subarrays of size $N^\text{R}_a$, then $\left\lceil \frac{N^\text{R}_a}{D_R} \right\rceil$ symbols are required to generate an isotropic combiner, since $K \geq N_a$ needs to be satisfied and every symbol can cover $D_\text{R}$ beams (one for each \ac{RF} chain). Thus, the minimum number of OFDM symbols required for a doubly isotropic combiner-pilot pair (with analog precoding) is $\left\lceil \frac{N^\text{R}_a}{D_\text{R}} \right\rceil  N_\text{T}$.\par

% An example of a $r_{mt}(\psi_\ell)$ designed in the described fashion can be seen on Fig.~\ref{fig:combiner_beams}, where the individual beams for each OFDM symbol can also be observed. The solid line black horizontal line corresponds to the sum of all beams, the dashed red line coincides with the aforementioned black line and is placed at the $D_R N^R_a N_\text{R}$ value (this is shown in Section \ref{sec:model_norm}).
% \begin{figure}[!htbp]
%     \centering
%     \includegraphics[width=0.4\linewidth]{combiner_beams_2rfchains_16subarray_24antennas.jpg}
%     \caption{Total combiner received power and individual combiner beams for each OFDM symbol. The considered scenario includes $D_R=2$, $N^R_a$, and $N_\text{R}=24$.}
%     \label{fig:combiner_beams}
% \end{figure}

\vspace{-0.2cm}
\subsection{Combiner-Pilot Design -- Hybrid Precoding}
Denote the $N_\text{T} \times 1$ steering vector to angle $\varsigma$ by $\mathbf{a}(\varsigma)$, also let $x_{nt}(\varsigma) = \sum^{D_\text{T}-1}_{d=0} \mathbf{a}^\text{T}(\varsigma) [\mathcal{P}]_{:dt} s_{ntd}$. Then, expanding
\begin{multline}
    x_{nt}(\varsigma) = \mathbf{a}^\text{T}(\varsigma) [\mathcal{P}]_{:0t} s_{nt0} + \mathbf{a}^\text{T}(\varsigma)[\mathcal{P}]_{:1t} s_{nt1} + \dots \\
    + \mathbf{a}^\text{T}(\varsigma)[\mathcal{P}]_{:(D_\text{T}-1)} s_{nt(D_\text{T}-1)},
\end{multline}
which has squared magnitude equal to 
\begin{multline}
    |x_{nt}(\varsigma)|^2 = \sum^{D_\text{T}-1}_{d=0} |\mathbf{a}^\text{T}(\varsigma)[\mathcal{P}]_{:dt}s_{ntd}|^2 \\ 
    + \!\!\!\!\!\!\!\!\!\!\!\!\! \!\!\!\! \! \sum_{\quad\: 0\leq d_1 < d_2 \leq D_\text{T}-1}   \!\!\! \!\!\!\!\!\!\!\!\!\!\!\!\! \Re{ \mathbf{a}^\text{T}(\varsigma)[\mathcal{P}]_{:d_1t}  [\mathcal{P}]^H_{:d_2t} \mathbf{a}^*(\varsigma)  s_{ntd_1}   s^*_{ntd_2}}.
\end{multline}
Summing over $n$ and $t$ to get the standard transmit isotropy term yields \vspace{-0.35cm}
\begin{multline} \label{eq:x_norm_expansion}
    \sum_{n,t} |x_{nt}(\varsigma)|^2 = \sum_{n,t} \sum^{D_\text{T}-1}_{d=0} |\mathbf{a}^\text{T}(\varsigma) [\mathcal{P}]_{:dt} s_{ntd}|^2 \\
    + \sum_{t}\!\!\!\!\!\! \!\!\!\!\!\!\!\!\!\! \sum_{\substack{\quad 0\leq d_1< d_2 \\ \quad \quad \quad\quad\quad \leq D_\text{T}-1} }  \!\!\!\!\!\!\!\!\!\!\!\!\!\!\!\! \Re{ \! \mathbf{a}^\text{T}(\varsigma)[\mathcal{P}]_{:d_1t}  [\mathcal{P}]^H_{:d_2t} \mathbf{a}^*(\varsigma)  
     \sum_n \! s_{ntd_1} s^*_{ntd_2} \!\! }.%[\mathcal{S}]^T_{:td_1} [\mathcal{S}]^*_{:td_2}}. % 
\end{multline}
%From the standard derivations of the isotropy condition (such as in \cite[Appendix~A]{ch_est_pinto}), one can see that, were the first term in the right hand side of (\ref{eq:x_norm_expansion}) by itself (with the precoders also chosen to cover an isotropic set, i.e., at least $N^T_a$ equally spaced transmit precoder angular frequencies), then the proposed pilot would be transmit isotropic. 
Comparing to the derivation of the isotropy condition in \cite[Appendix~A]{ch_est_pinto}, we can see that (\ref{eq:x_norm_expansion}) would be transmit isotropic if it contained only its first right-hand side term with the precoders chosen to cover an isotropic set (i.e., at least $N^\text{T}_a$ equally spaced transmit precoder angular frequencies).
However, the crossterms prevent the pilot from satisfying the isotropy condition. It is possible to eliminate the crossterms by choosing symbols such that $\sum_n s_{ntd_1} s^*_{ntd_2} = 0$ for all $t$ and distinct $(d_1,d_2)$ pairs. This is equivalent to stating that the pilot vectors should be orthogonal over $n$ for different RF chains. With ideal hardware, any orthogonal dictionary may be used, e.g., Hadamard, \ac{DFT}, \ac{ZC}. However, for \ac{OFDM} with impaired hardware, using \ac{DFT} sequences leads to impulse-like waveforms in the time domain, which are severely affected by amplifier distortion due to poor \ac{PAPR}. Therefore, we use \ac{ZC} sequences throughout the remainder of this paper.

Effectively, the minimum number of OFDM symbols required for generating a doubly isotropic combiner-pilot pair is $\left\lceil \frac{N^\text{R}_a}{D_\text{R}} \right\rceil  \left\lceil \frac{N^\text{T}_a}{D_\text{T}} \right\rceil$. The minimum pilot sequence length for hybrid beamforming can be substantially smaller than the one for analog beamforming (equal to $N_\text{T} N_\text{R}$), which may allow for significant reduction of pilot overhead in sensing applications when compared to single-beam beamsweeping. Naturally, using subarrays instead of the full array leads to angular resolution loss.\par

\vspace{-0.5cm}
\section{Multiple Start SAGE} \label{sec:ms_sage}

We adopt the approach from \cite{ch_est_pinto} in which paths are progressively introduced to the model up to convergence or the satisfaction of a global threshold. Due to the nonconvex nature of the optimization problem, local descent algorithms like \ac{SAGE} -- which are a subset of \ac{MM} methods -- are sensitive to initialization conditions. This issue has been addressed in \cite{ch_est_pinto} using a reinitialization strategy, in which paths that provide insufficient improvement\footnote{In \cite{ch_est_pinto}, the magnitude of the improvement is determined by a threshold, which is computed from the properties of the \ac{GLRT} statistic. If the improvement could be explained by white noise within an upper bound on the probability of false alarm, then the improvement is judged insufficient and the path is reinitialized.} to the value of the objective function are reinitialized randomly. This is repeated until either a suitable initialization point is found, upon which the algorithm proceeds with the updates, or until a maximum number of reinitializations is reached, upon which the algorithm halts. The work in \cite{ch_est_pinto} considers exclusively a fully digital combiner and a fully analog precoder. However, it turns out that using a reinitialization threshold derived in the same fashion as \cite{ch_est_pinto} leads to an algorithm that is very sensitive to model mismatch, which exists due to Assumption~\ref{assumption:noise_modeling}.
\textcolor{blue}{
We first describe the considered algorithm below and subsequently study its computational complexity.}

\textcolor{blue}{\subsection{Algorithm Description}}
To address the issue of sensitivity to initialization points \textcolor{blue}{in a robust manner,} we propose a multiple start heuristic \textcolor{blue}{as a substitute for the reinitialization approach from \cite{ch_est_pinto}}. Whenever a new path is introduced, a total of $K_\text{init}$ random initialization points are attempted, each of them is subjected to the coordinate descent procedure described in Algorithm~\ref{alg:inner}. The point that provides the largest improvement to the objective function is selected as the definitive initialization point for the new path. The multiple start approach is more expensive in terms of computational resources. By sheer extra computational effort, the multiple start heuristic necessarily outperforms the reinitialization approach if the maximum number of reinitializations is the same as the number of initialization points. In the worst case, reducing the number of initialization points leads to model order overestimation (adding negligible magnitude paths), with no significant impact on the resulting channel tensor error. The path initialization method is described in Algorithm~\ref{alg:path_init}. Many properties of the \ac{MS} \ac{SAGE} algorithm (such as computational complexity, convergence, accuracy, etc.) can be understood as conceptual extensions of \cite{ch_est_pinto}, thus we omit it in this work. Further theoretical exploration of the algorithm itself, as well as improving on model selection methods under hardware impairments, is left for future work.

\begin{algorithm}[!htbp]
\caption{Inner loop: path optimization routine}
\label{alg:inner}
\begin{algorithmic}[1]
    \Procedure{UpdatePath}{$\Tilde{\mathcal{Y}}_\ell$, $\mathcal{X}$, $\ell$, $\boldsymbol{\xi}_{\ell}$}       %\Comment{This is a test}
    \For{$\text{it}_\text{in} = 1,\,\dots,\,\text{it}^\text{max}_\text{in}$} 
        \State Compute $\frac{\partial h(\psi_{\ell})}{\partial \psi_{\ell}}$ and get real roots;
        \State Set $\psi_{\ell}$ to the root with highest $h(\psi_{\ell})$;
        \State Do the same for $\omega_{1\ell}$, $\omega_{2\ell}$, and $\varsigma_{\ell}$ (use numerical line search on $\varsigma_\ell$ if \ac{Tx} impairments are known);\par
        
    \EndFor 
    \State Compute $b_{\ell}$ using (\ref{eq:b_opt_hybrid});
    \State \Return $\boldsymbol{\xi}_{\ell}$; 
\EndProcedure
\end{algorithmic}
\end{algorithm}

\begin{algorithm}[!htbp]
\caption{Path initialization: multiple start heuristic}
\label{alg:path_init}
\begin{algorithmic}[1]
    \Procedure{PathInit}{$\Tilde{\mathcal{Y}}_\ell$, $\mathcal{X}$, $K_\text{init}$}       %\Comment{This is a test}
    \State $g_\text{best} = 0$;
    \For{$k = 1,\,\dots,\,K_\text{init}$} 
        \State Initialize $\boldsymbol{\xi}_\ell$ randomly;
        \State $\boldsymbol{\xi}_\ell = $UpdatePath($\Tilde{\mathcal{Y}}_\ell$, $\mathcal{X}$, $\ell$, $\boldsymbol{\xi}_{\ell}$);
        \If{$g(\boldsymbol{\xi}_\ell) > g_\text{best}$}
            \State $g_\text{best} = g(\boldsymbol{\xi}_\ell)$; $\boldsymbol{\xi}_\text{best} = \boldsymbol{\xi}_\ell$;
        \EndIf
    \EndFor 
    \Return $\boldsymbol{\xi}_\text{best}$
\EndProcedure
\end{algorithmic}
\end{algorithm}
%\FloatBarrier

Once a path is initialized, it is updated in a cyclic fashion along with the other paths initialized earlier. If a path provides a relative improvement to the objective function smaller than a threshold $\epsilon_\text{steady}$, then it is no longer updated within this instance of the loop. The path update loop is described in Algorithm~\ref{alg:middle}.

\begin{algorithm}[!htbp]
\caption{Middle loop: path update loop}
\label{alg:middle}
\begin{algorithmic}[1]
    \Procedure{UpdateLoop}{$\mathcal{Y}$, $\mathcal{X}$, $\boldsymbol{\xi}$, $L$}       % , $ , $L_\text{TW}$ \text{it}_\text{mid}$ $\epsilon_\text{init}$, $\epsilon_\text{steady}$
    \State $\mathcal{L}_\text{steady} = \{\}$;
        \State \textcolor{blue}{$L_\text{order} = [L,\, 1,\,\dots,\, L-1]$;}
    \For{$\text{it}_\text{mid}=1,\,\dots,\,\text{it}^\text{max}_\text{mid}$}
        \If{$\mathcal{L}_\text{steady} \!=\! \{ \ell\!:\!\ell\!=\!L_\text{order}(k), k=1,\dots,L \}$}
            \State Break; \Comment{Stop loop if all paths are stable}
        \EndIf 
        \State $k_\text{path}=1$;
        \While{$k_\text{path}\leq L$}
            
            \State $\ell = L_\text{order}(k_\text{path})$;
            \If{$\ell \in \mathcal{L}_\text{steady}$} Skip iteration;
            \EndIf
            \State Compute $\Tilde{\mathcal{Y}}_\ell$, and $f_\text{old} = f(\boldsymbol{\xi})$;
            \State $\boldsymbol{\xi}_{\ell}$ = UpdatePath($\Tilde{\mathcal{Y}}_\ell$, $\mathcal{X}$, $\ell$, $\boldsymbol{\xi}_{\ell}$);
            \If{$f(\boldsymbol{\xi})< \epsilon$} Stop execution; \Comment{Global threshold reached}
            \EndIf
            %\State $\hat{\epsilon}_\text{init} = \min(\epsilon_\text{init}, \hat{f}(\boldsymbol{\xi}_\ell)-\hat{f}_\text{thld})$ \label{algline:reinit_thld_min}
            \If{($f_\text{old} - f(\boldsymbol{\xi}))/f_\text{old} \leq \epsilon_\text{steady}$}
                \State $\mathcal{L}_\text{steady} = \mathcal{L}_\text{steady} \cup \ell$;
                %\State $\Delta f_\ell  = \Delta f_\ell + \hat{f}_\text{old} - \hat{f}(\boldsymbol{\xi}_\ell)$
            \EndIf
            \State $k_\text{path}=k_\text{path}+1$;
        \EndWhile
    \EndFor   
    \Return $\boldsymbol{\xi}$
    \EndProcedure
\end{algorithmic}
\end{algorithm}

The full algorithm progressively introduces new paths, which are initialized using Algorithm~\ref{alg:path_init}, and updates the paths in the active path pool cyclically using Algorithm~\ref{alg:middle}. A high level description of the full algorithm is shown in Algorithm~\ref{alg:outer}. As in \cite[Sec. V]{ch_est_pinto}, the algorithm halts once a global threshold \textcolor{blue}{$\epsilon$ computed from $N^Z_0$} is reached. If the objective function is smaller than the threshold, then it is likely that the residual data (i.e., the data after subtracting all inferred paths from it) is generated exclusively by white noise with \textcolor{blue}{confidence level}~$\delta$. \textcolor{blue}{However, differently from \cite{ch_est_pinto}, the \ac{MS}-\ac{SAGE} algorithm also stops if a proposed path provides insufficient improvement. Improvement is considered insufficient if the objective does not decrease by more than
\begin{equation} \label{eq:path_acc_thld}
    \epsilon_\text{path} \;=\; 2\log N' + 3\log\log N' + \log\!\left(\frac{448\pi^3}{81\,p_{fa}^2}\right),
\end{equation}
with $N' = N_a^{R}N_a^{T}N_SN_C$ being the number of resolvable cells of the 4-D search and $p_{fa}$ the per-path false-alarm probability. The expression in (\ref{eq:path_acc_thld}) is equivalent to the path reinitialization threshold in \cite{ch_est_pinto}}. \textcolor{blue}{A polishing pass is performed after the global threshold is reached to help ensure convergence was reached.}

\begin{algorithm}[!htbp]
\caption{Main loop}
\label{alg:outer}
\begin{algorithmic}[1]
    \Procedure{Main}{$\mathcal{Y}$, $\mathcal{X}$}       %\Comment{This is a test}
    \For{$L = 1,\,\dots,\,L_\text{max}$} 
        \State Compute $\Tilde{\mathcal{Y}}_L$ using (\ref{eq:y_tilde})
        \State $\boldsymbol{\xi}_L$ = PathInit($\Tilde{\mathcal{Y}}_L$, 
        $\mathcal{X}$, $K_\text{init}$)
        \textcolor{blue}{
        \If{$f(\boldsymbol{\xi}_{1:L-1}) - f(\boldsymbol{\xi}_{1:L})\leq\epsilon_\text{path}$}
            \State Break; \Comment{Stop if insufficient improvement}
        \EndIf
        \State $\boldsymbol{\xi}$ = UpdateLoop($\mathcal{Y}$, $\mathcal{X}$, $\boldsymbol{\xi}$, $L$);
        \If{global threshold reached}
            \State Break;
        \EndIf}      
    \EndFor \par
    \State \pushcode[-0]\textcolor{blue}{UpdateLoop($\mathcal{Y}$, $\mathcal{X}$, $\boldsymbol{\xi}$, $L$) with disabled global stopping conditions;} \Comment{Polishing pass for convergence}\par
    \Return $\boldsymbol{\xi}$, $L$
\EndProcedure
\end{algorithmic}
\end{algorithm}

\vspace{-1cm}
\textcolor{blue}{
\subsection{Computational Complexity}
We provide a brief analysis of the computational complexity of the \ac{MS} \ac{SAGE} algorithm. Each inner iteration updates the four parameters of a path with each update consisting of a tensor build and contraction, $\mathcal O(D_RN_SN_CN_{\rm par})$, plus companion-matrix root finding on the derivative polynomial, $\mathcal O(N_{\rm par}^3)$, with $N_{\rm par}\in\{N_C,N_T,N_S,N_R\}$. Multiple start initialization costs $K_{\rm init}\,it_{\rm in}^{\max}$ inner iterations per added path. Middle loop iteration count depends on convergence properties of each path, but we can upper bound it by $\mathcal{O}(\text{it}^\text{max}_\text{mid} L_\text{est})$, where $L_\text{est}$ corresponds to the estimate of $L$ when \textit{UpdateLoop} is called. Since $L_\text{est}$ ideally increases from 1 to $L$, the number of middle loop iterations across the whole algorithm execution is approximately $\mathcal{O}(L(L+1)\text{it}^\text{max}_\text{mid}/2)$. The total complexity is then given by the initializations and total path updates, namely
\begin{equation}
    \mathcal{C}_\text{tot} = \left(L K_{\rm init}\,\text{it}_{\rm in}^{\max} + \frac{L(L+1)\text{it}^\text{max}_\text{mid}}{2}\right) \mathcal{C}_\text{inner},
\end{equation}
where $\mathcal{C}_\text{inner}$ is the complexity of an inner loop iteration.
}
\vspace{-0.4cm}
\section{Numerical Results} \label{sec:numres}

In this section, we numerically evaluate the parametric channel estimation, sensing (detection and estimation), and communications performances of different hybrid beamforming architectures affected by hardware impairments. Concurrently, we also evaluate the power efficiency of such architectures and study the tradeoff between performance and power consumption. The results are achieved from \ac{MC} simulations over varying \ac{ADC} resolution for multiple beamformer architectures. A total of 256 samples are computed for each data point. 

\vspace{-0.3cm}
\subsection{Simulation Scenario}
The simulations concern a bistatic sensing scenario between two devices with symmetrical precoding and combining architectures. This means that both \acp{BS} have the same number of antennas, \ac{RF} chains, as well as subarrays of the same size. This is done as a simplifying approach to avoid sweeping over all possible combinations of \ac{Tx} and \ac{Rx} architectures, which would make data presentation very challenging, as well as reducing the interpretability of the results. The simulation consists of two frames: a pilot/sensing frame and a communications frame. The pilot frame is used for \ac{PCE} and extracting sensing information, while the communications step uses the channel estimate for equalization and delivers the data payload. We assume that the \ac{Rx} does not have a model of the \ac{Tx} impairments, such as described in Item \ref{item:not_known_impairments} of Section \ref{subsubsec:sigma_update}. \par

\vspace{-0.5cm}
\subsection{Data Generation}
\vspace{-0.1cm}
\subsubsection{Channel Generation}
The simulation aims to represent a transmission in the upper-midband with a carrier of 7.125~GHz. Similar to \cite{ch_est_pinto}, the channel is generated as the sum of $L$ rank-1 tensors with angular frequencies sampled from uniform distributions, i.e., $(\omega_{1\ell},\omega_{2\ell},\psi_\ell,\varsigma_\ell)\sim\mathcal{U}^4(-\pi,\pi)$. The phases of the path coefficients are also uniformly distributed, $\angle b_\ell \sim \mathcal{U}(-\pi,\pi)$. The path coefficient magnitudes are sampled from a positive support \ac{PDF}, a Rician distribution with noncentrality parameter $10^{-6}$ and scale parameter $5\cdot10^{-6}$ has been chosen in this case. The largest magnitude path also receives a linear gain of $a_\text{LOS}>1$, representing the dominant \ac{LOS} component of the channel. In the presented simulations, we consider $a_\text{LOS}=10$. A total of $L=8$ paths are generated, \textcolor{blue}{ordered $|b_1|\geq|b_2|\geq\dots\geq|b_L|$}, with the lowest magnitude path representing a sensing target. Both the \ac{Tx} and \ac{Rx} have $N_\text{T} = N_\text{R} = 8$ antennas, which are divided into \mbox{$D_\text{T} = D_\text{R} \in \{ 1,2,4,8 \}$} subarrays with no overlapping antennas between arrays (i.e., sub-panel beamformer architecture). The parameter generation is summarized in Table~\ref{tab:param_gen}. The ground truth channel tensor is computed from the sampled parameters according to (\ref{eq:channel_tensor}).\par %The lowest magnitude component receives a linear decrement of $a_\text{S}<1$, representing the smaller magnitude of sensing target reflections. \par

\begin{table}[!htbp]
    \centering
    \caption{Summary of the simulation parameter generation.}
    \begin{tabular}{c|c}
     Parameter(s)    &  Relevant Information\\ \hline
     $\omega_{1\ell},\omega_{2\ell},\psi_\ell,\varsigma_\ell,$ and $\angle b_\ell$    & $\mathcal{U}(-\pi,\pi)$ distributed \\
     Path magnitudes $|b_\ell|$    & Rice($10^{-6},5\cdot10^{-6}$) distributed \\
     Largest generated $|b_\ell|$ & Multiplied by $a_\text{LOS}=10$ \\
     Number of paths $L$ & Set to $L=8$ \\
     Lowest magnitude path & Represents the sensing target \\
     Number of \ac{Tx} and \ac{Rx} antennas & $N_\text{T} = N_\text{R} = 8$ \\
     Number of \ac{RF} chains & $D_\text{T} = D_\text{R} \in \{ 1,2,4,8 \}$ \\
     Beamforming topology & Sub-panel
     
    \end{tabular}
    \label{tab:param_gen}
\end{table}

\subsubsection{Pilot and Communications Frames}
The transmitted pilot sequence consists of $N_\text{S}=64$ \ac{OFDM} symbols with $N_\text{C}=31$ subcarriers. The subcarrier data of each stream consists of randomly sampled \ac{ZC} sequences, orthogonal between different streams. The pilot frame is a doubly isotropic pilot-combiner pair. The communications frame has the same resources of the pilot frame, i.e., $N_\text{S}=64$ and $N_\text{C}=31$, but the precoder and combiner directions are beamformed towards the estimated highest power path, that is, the precoder is matched to $\hat{\varsigma}_1$ and the combiner to $\hat{\psi}_1$. The communications data is a single stream of independent \ac{QPSK} symbols. Both the pilot and communications frames are transmitted at a total power of $30$~dBm, which is high enough such that the performance is mostly limited by hardware impairments and not thermal noise.

\subsubsection{Hardware Impairments}
Regarding quantizers, for the \ac{ADC}, both \ac{LM} optimal quantizers and uniform quantizers are considered as defined in Section \ref{subsec:quantizers}. For the \ac{LM} quantizer, the quantization regions and representation points are computed from a dataset of 50 pilot transmissions. The backoff factors are set to $\nu_\text{R} = \nu_\text{T} = 4$. The \ac{Rx} backoff factor during the communications frame is reduced to $\nu_R = 3$, which improves communications performance for the uniform \ac{ADC} scenario without significant effect on the \ac{LM} quantizer case. \par

Since the optimal \ac{ADC} is trained from a dataset of ideal isotropic pilots, the statistics of the received data will be different for the communications frame. Assuming unit path gain, the energy received from a path when using the proposed doubly isotropic pilot combiner structure will be equal to 
\begin{equation} \label{eq:model_norm}
    \sum_{ntu} |r_{mt}(\psi)|^2 |x_{nt}(\varsigma)|^2 = D_\text{R} N^\text{R}_a N_\text{S} \rho,
\end{equation}
where $\rho$ is the transmitted power. The quantity in (\ref{eq:model_norm}) is also called the model square norm and its value is derived in Appendix \ref{appendix:model_norm_deriv}. If the parameters of the path are perfectly known, then the subarray beamformer and combiner lead to a model order gain of $a_\text{comms} = N^\text{R}_a N_\text{T}$. The gain value is derived in Appendix \ref{appendix:bf_gain}.
Although the signal statistics differ, the same \ac{ADC} parameters from the pilot frame may be used in the communications frame without significant performance loss if the quantization regions and representation points are scaled by $\sqrt{a_\text{comms}}$. This is done to avoid recomputing an optimal quantizer for every frame, with the consequence of foregoing quantizer optimality in the square error sense.\par

The \ac{DAC} in the \ac{Tx} is a uniform \ac{DAC} computed from the backoff factor as explained in Section \ref{subsec:quantizers}. The \ac{DAC} resolution is kept constant at 10 bits. This resolution is chosen such as to still present noticeable impairment effect when the \acp{ADC} have high resolutions, but while still maintaining acceptable \ac{EVM} performance in the communications frame.

Regarding the amplifiers, the \ac{LNA} is known to operate in more linear fashion than the \ac{PA} (which should operate close to saturation for power efficiency reasons). Therefore we set the nonlinearity parameters as $\kappa_\text{LNA} = 10$ and $\kappa_\text{PA} = 3$. The nonlinearity parameters of the amplifiers are kept constant throughout all simulations. 

\subsubsection{Thermal Noise Modeling}
For the thermal noise, we consider a power spectral density of $\eta = -173~$dBm/Hz, which results from the Johnson-Nyquist noise expression $\eta = k_B T$, where $k_B$ is the Boltzmann constant, using a temperature of $T=290~K$. The noise is enhanced by a noise figure of $F = 10~$dB. For the bandwidth, we consider a subcarrier spacing of $f_{sc} = 30$~kHz, inspired by the \ac{5G} standard numerology $\mu=1$ as detailed in TS 38.211 Release~18 \cite{3gpp.38.211}. Thus, $N_0 = \eta f_{sc} F$ (where $F$ is converted from dB to linear). Estimating the hardware impairment noise is a non-trivial matter which we consider beyond the scope of this work, but nonetheless a prospective direction for future work. For simplicity, using Assumption \ref{assumption:noise_modeling}, we compute the average ground truth impairment square error 
\textcolor{blue}{
\begin{equation} \label{eq:imp_err_est}
    N_0^{z}= N_0 + \hat{N}^z_0, \; \hat{N}^z_0 =  \frac{1}{N_CN_SD_R}\left\|\Bar{\mathcal Y}-\mathcal W-c\,\mathcal Y\right\|_F^2, 
\end{equation}
where $\mathcal Y$ is the impairment-free signal (with no thermal noise, as understood in the context of Assumption~\ref{assumption:noise_modeling}), $\mathcal W$ the thermal-noise realization, and $c=\langle\mathcal Y,\bar{\mathcal Y}-\mathcal W\rangle/\lVert\mathcal Y\rVert^2_F$ the least-squares fit of $\bar{\mathcal Y}-\mathcal W$ onto $\mathcal Y$.} 
\textcolor{blue}{The \ac{SAGE} estimator fits the ideal signal model to the received data. Subtracting the least-squares fit $c \, \mathcal Y$ leaves in $\hat N_0^{z}$ only what the signal model cannot reproduce, since the estimated path coefficients absorb any common complex scaling of the signal. This expression allows us to model the impacts of \ac{Tx} and \ac{Rx} impairments. We use $N_0^{z}$ as an estimate for the total noise variance.}

%\textcolor{blue}{Since the path coefficients $b_\ell$ are estimated, an impairment that scales the received pilot by a common complex factor is indistinguishable from those coefficients and does not degrade the estimate; subtracting the least-squares fit $c \, \mathcal Y$ leaves in $\hat N_0^{z}$ only what the signal model cannot reproduce. This expression allows us to model the impacts of \ac{Tx} and \ac{Rx} impairments. We use $N_0^{z}$ as an estimate for the total noise variance.}

\vspace{-0.4cm}
\subsection{Channel Estimator Parameters}
The \ac{MS}-\ac{SAGE} algorithm is executed with a maximum total of 3000 inner iterations. An inner iteration is defined as an execution of the inner loop of the UpdatePath routine of Algorithm \ref{alg:inner}. The number of initialization points is set to $K_\text{init} = 20$. The algorithm is executed with $\text{it}^\text{max}_\text{in} = 2$, $\text{it}^\text{max}_\text{mid} = 100$, $L_\text{max} = 20$, and no trailing window. The steadiness threshold is set to $\epsilon_\text{steady}=10^{-16}$. \textcolor{blue}{The global threshold parameter is set to $\delta = 0.95$ and the per-path stopping parameter is set to $p_{fa}=10^{-8}$.}

\vspace{-0.4cm}
\subsection{Performance Metrics} \label{subsec:perf_metrics}
The parametric channel estimation performance is evaluated by assessing the relative error of the channel tensor estimate given by $\| \mathcal{H} - \hat{\mathcal{H}} \|_F/\|\mathcal{H}\|_F$ as well as the estimated model order \textcolor{blue}{$\hat{L}$, i.e., the estimate of} the number of paths $L$. It should be noted that meaningful differences in resolution between beamformer topologies are implicitly considered by this channel error metric. This happens because any resolution loss that degrades path detection performance should also lead to a consequent increase in channel estimation error.\par

The sensing performance is assessed by assigning the lowest magnitude path to represent a sensing target of interest, the detection and estimation of this path then indicates the sensing performance of the system. We remark that it is not straightforward to strictly state whether a path is detected or not. To formalize the notion of a path detection, we could check whether there exists an estimated path within some parametric distance of the ground truth path. \textcolor{blue}{This requires establishing some form of assignment between the detected paths and ground truth paths. First, we perform a greedy minimum parameter distance assignment as follows
\begin{gather} 
    k^{\star}_\ell=\arg\min_{k\in\{1,\dots,\hat L\}}\frac{1}{4\pi}\bigl\lVert\boldsymbol\chi_{\ell}-\hat{\boldsymbol\chi}_{k}\bigr\rVert_{S} \\ \label{eq:path_param_err}
    e^\ell_{p}=\frac{\bigl\| p_{\ell}-\hat p_{k^{\star}_\ell} \bigr\|_S}{\pi},\; p\in\{\omega_{1},\omega_{2},\psi,\varsigma\},\; e^\ell_{p}\in[0,1]
\end{gather}
where $\boldsymbol{\chi}_\ell = \{\omega_{1\ell}, \omega_{2\ell}, \psi_\ell, \varsigma_\ell\}$ denotes the ground truth parameters of path $\ell$, and $\hat{\boldsymbol\chi}_k$ denotes the estimated parameters of path $k\in\{1,\dots,\hat L\}$. Also, $\|\boldsymbol\chi\|_S=\sum_p|\angle e^{j\chi_p}|$ is the \mbox{$l_1$-norm} on the torus of appropriate dimension; for scalars, it is the magnitude of the value wrapped to $(-\pi,\pi]$, hence bounded by $\pi$.
Therefore, beyond the individual parameter errors, we define the $\ell$th \ac{PPE} as a path estimation metric
\begin{equation}
    \ell\text{th-PPE} = \min_k \frac{1}{4\pi} \| \boldsymbol{\chi}_\ell - \hat{\boldsymbol{\chi}}_k \|_S = \frac14\sum_p e^\ell_p,
\end{equation}
 The $\ell$th-PPE can be understood as the error of the closest path to path $\ell$ normalized by $4\pi$, which is the maximum possible distance on the 4-torus. This metric is biased towards the worst parameter estimate. However, if we interpret sensing performance to be good when the estimation of every parameter is good, then the $\ell\text{th-PPE}$ metric is an indicator of whether the estimation is accurate in every sensing parameter.}

The communications performance is assessed by transmitting a single data stream through the same channel and evaluating the resulting \ac{EVM}. The inverse channel estimate is used as a zero-forcing equalizer. \par

The power consumption aspects and their relation to the channel estimation, sensing, and communications performance are assessed by the inverse of the product between the respective metrics---i.e., relative channel error, $\ell$th-PPE, and \ac{EVM}---and the power consumption values. For example, the channel estimation power efficiency is given by 
\begin{equation}
    \text{PE}_\text{ch} = \frac{\|\mathcal{H}\|_F}{\| \mathcal{H} - \hat{\mathcal{H}} \|_F} \frac{1}{P_\text{tot}(D,\! N^\text{bits}_\text{DAC}, \! N^\text{bits}_\text{ADC})} \text{  (error reduction)/W},
\end{equation}
which indicates by how much each watt of power is able to decrease the relative channel estimation error on average. For example, a value of $\text{PE}_\text{ch}=2$ indicates that, on average, each watt of total consumed power led to a reduction of 50\% in relative channel estimation error. Power efficiency is similarly defined for the other metrics, namely $\text{PE}_\text{sens}$ for the sensing efficiency and $\text{PE}_\text{com}$ for the communications efficiency.

\vspace{-0.3cm}
\subsection{Channel Est., Sensing, and Communications Performance} 
Because the distributions of performance metrics prove to be asymmetric, we present both the mean (solid line) and the median (dashed line) to have a more complete performance characterization. Each figure presents the results for \ac{LM} and uniform quantizers. \par

\begin{figure}[!htbp]
    \centering
    \includegraphics[width=1\linewidth]{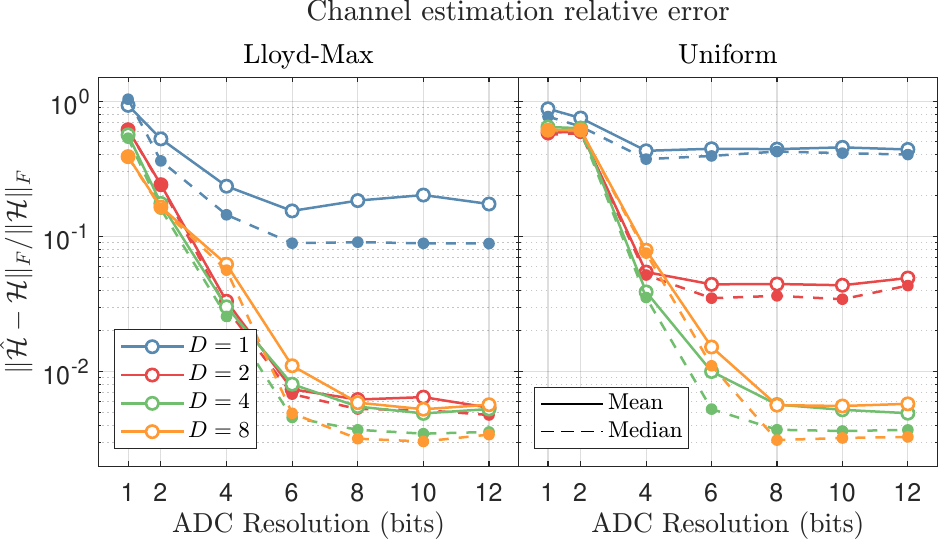}
    \caption{Relative channel estimation error, optimal (left) and uniform (right) \ac{ADC} for different numbers of RF chains.}
    \label{fig:ch_est_err}
\end{figure}

Fig.~\ref{fig:ch_est_err} shows the relative channel estimation error for four different transceiver architectures, indicated by the number of \ac{RF} chains $D$, and for varying \ac{ADC} resolution, given in bits. It can be seen that, as expected, improving the \ac{ADC} resolution leads to better channel estimation performance for every architecture, for both the \ac{LM} and uniform quantizers. It is also intuitive that increasing the number of \ac{RF} chains leads to improved estimation performance. This can also be observed in Fig.~\ref{fig:ch_est_err}, especially for resolutions above 8 bits. Notice that, after a particular \ac{ADC} resolution, the uniform \ac{ADC} curves reach a performance floor caused by errors at the ends-of-scale of the quantizers. To avoid such performance floors, the \ac{ADC} range should grow as the resolution increases.\par

\begin{figure}[!htbp]
    \centering
    \includegraphics[width=1\linewidth]{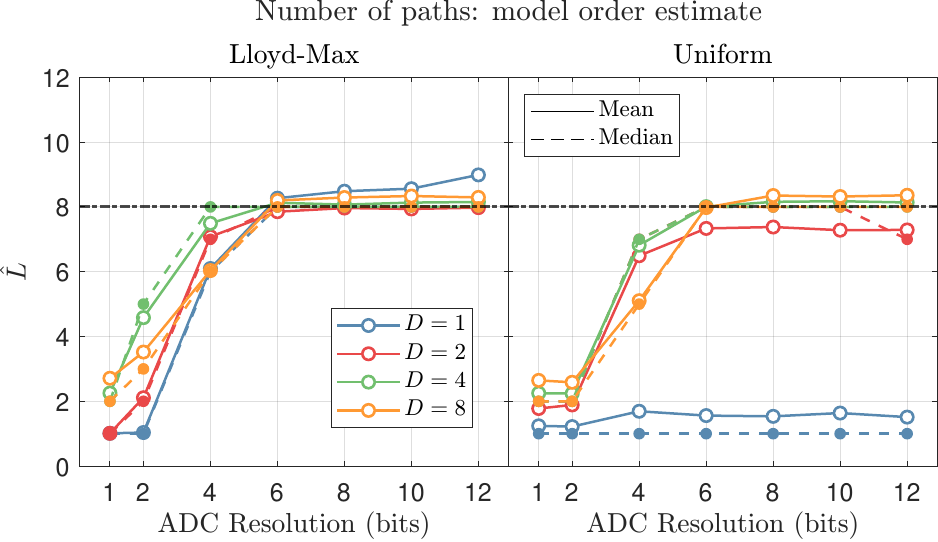}
    \caption{Model order estimates for different numbers of RF chains. Ground truth $L=8$ is indicated by dash-dotted black line.}
    \label{fig:L_err}
\end{figure}

\textcolor{blue}{The model order estimation behavior is shown in Fig.~\ref{fig:L_err}.} Model order estimation with hardware impairments is more complicated than in the ideal case considered in \cite{ch_est_pinto}. This is primarily due to the difficulty of modeling impairment noise in a simple enough fashion. The use of Assumption \ref{assumption:noise_modeling} leads to model order estimation error \textcolor{blue}{in the high \ac{ADC} resolution regime due to model mismatch. It can be seen that, for the \ac{LM} quantizer, there is a slight model order overestimation for all values of $D$. For the uniform quantizer, $D=1$ severely underestimates $L$. This happens because the estimation algorithm quickly reaches the global stopping threshold due to the high quantization error.}

%as the \ac{ADC} resolution increases. Because the global stopping criterion of the algorithm is a function of the sum of  thermal and impairment noise components, increasing the \ac{ADC} resolution (thus decreasing the quantization error power) allows the algorithm to execute for more iterations than would be necessary. This behavior can be observed in Fig.~\ref{fig:L_err}. 

\begin{figure}[!htbp]
    \centering
    \includegraphics[width=1\linewidth]{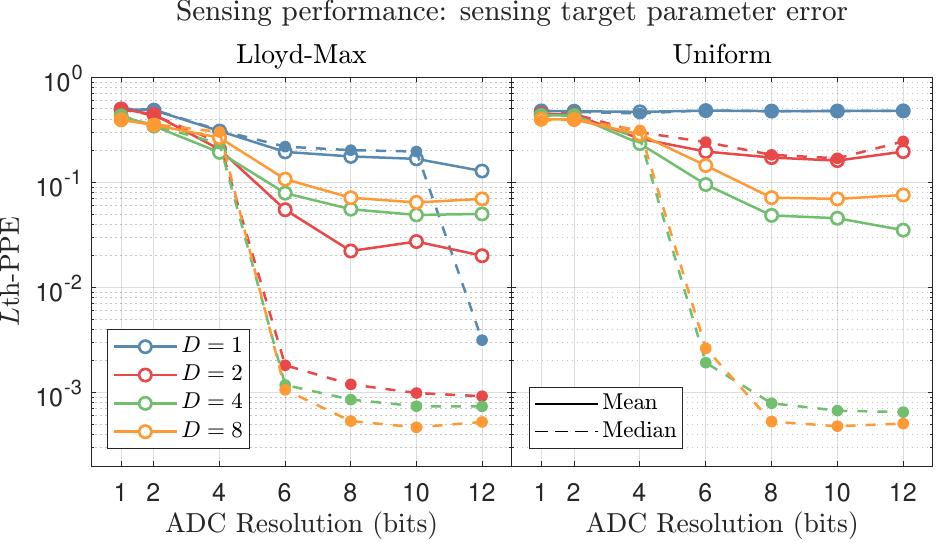}
    \caption{Sensing target \ac{PPE} for different numbers of RF chains.}
    \label{fig:sensing_ppe}
\end{figure}

The detection and estimation performances of a sensing target are presented in Fig.~\ref{fig:sensing_ppe}. As expected, improving the \ac{ADC} resolution noticeably improves the sensing performance, since the low magnitude sensing path is no longer overpowered by hardware impairment noise. Additionally, increasing the number of \ac{RF} chains seems to lead to better detection and estimation of the target. However, for optimal \acp{ADC}, in the median sense, there are diminishing returns on increasing $D$. Additionally, employing more \ac{RF} chains leads to increased robustness against quantization noise, as visible in Fig.~\ref{fig:sensing_ppe}.\par

\begin{figure*}[!htbp]
    \centering
    \includegraphics[width=1\linewidth]{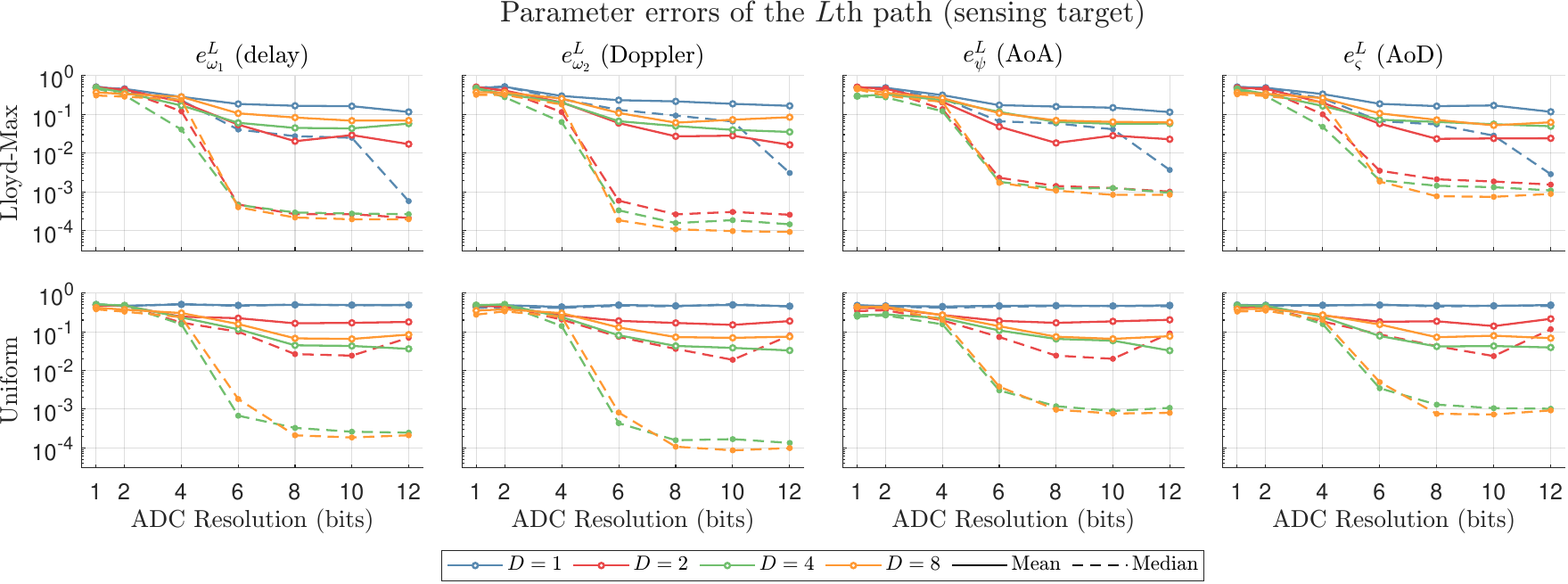}
    \caption{\textcolor{blue}{Individual parameter errors for the $L$th path (sensing target) for $D\in\{1,2,4,8\}$.}}
    \label{fig:per_param}
\end{figure*}

\textcolor{blue}{The $L\text{th-PPE}$ sensing performance metric gives a compounded number over all the sensing parameters. Thus, to assess the sensing performance over each parameter we compute $e^L_{p}$ as defined in (\ref{eq:path_param_err}) and display the achieved results in Fig.~\ref{fig:per_param}. In general, it can be seen that the individual error curves have similar shape to their corresponding curve in Fig.~\ref{fig:sensing_ppe}, which indicates that the $L\text{th-PPE}$ is a reliable indicator of relative sensing performance. Therefore, we can draw similar conclusions from both figures. In absolute terms, the parameters associated with larger dimensions (namely, $\omega_1$ and $\omega_2$) have smaller error, as expected.}

\begin{figure}[!htbp]
    \centering
    \includegraphics[width=1\linewidth]{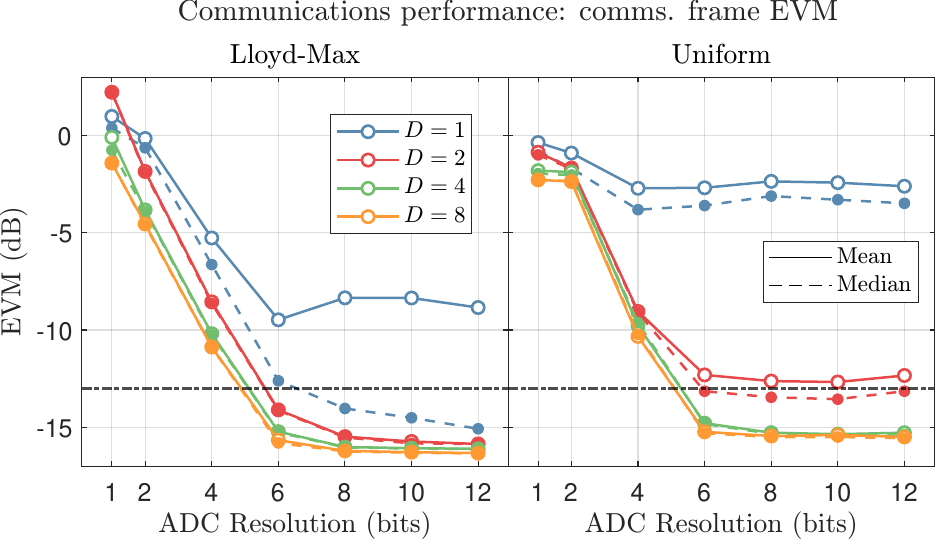}
    \caption{Communications frame \ac{EVM} for different numbers of RF chains. The dash-dotted line indicates the $-13$~dB operating point.}
    \label{fig:comms_evm}
\end{figure}

In Fig.~\ref{fig:comms_evm}, we present the \ac{EVM} of the symbols in the communications frame. In the figure, for reference, we also indicate a typical \ac{EVM} operating point of 5\% (i.e., $-13$~dB), which is roughly based on the values presented in \cite{evm_operating_point}. 
As expected, raising the \ac{ADC} resolution reduces the \ac{EVM}. 
However, the \ac{EVM} eventually reaches a performance floor as the \ac{ADC} resolution increases due to the effects of the remaining hardware impairments, such as finite \ac{DAC} resolution.%and amplifier nonlinearities.

As expected, overall channel estimation performance positively correlates with sensing and communications performances, as evidenced by the results shown in Fig.~\ref{fig:ch_est_err}, Fig.~\ref{fig:sensing_ppe}, and Fig.~\ref{fig:comms_evm}. In the sensing aspect, this can be justified by the fact that improving the channel estimates requires including all true paths with accurate parameter estimation. Therefore, as channel estimation improves, it is likely that sensing performance also improves, since detection and accurate estimation of the sensing target and its parameters is more likely. From the communications perspective, a better channel estimate has two main effects: improved equalization performance and symbol recovery, and also more accurate estimation of the \ac{LOS} path parameters for precoding and combining during the communications frame.\par

\begin{figure*}[htbp!] 
    \begin{subfigure}{0.8\columnwidth}
        \centering
        \includegraphics[width=0.735\columnwidth]{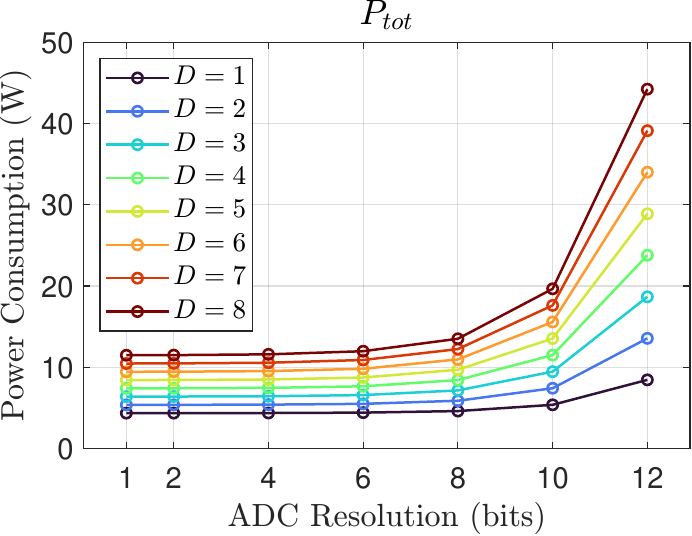}
        \caption{Power consumption for varying $D$ and \ac{ADC} resolution.}
        \label{subfig:pwr_cons}
    \end{subfigure} \hfill
    \begin{subfigure}{1.2\columnwidth}
        \centering
        \includegraphics[width=1.\columnwidth]{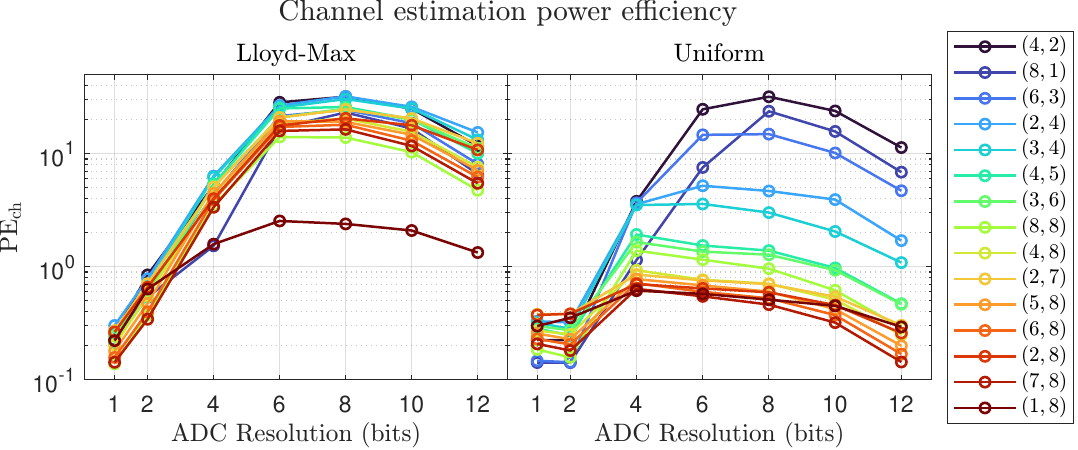}
        \caption{Channel estimation error and power tradeoff.}
        \label{subfig:ch_est_pwr}
    \end{subfigure} \hfill
    \begin{subfigure}{1\columnwidth}
        \centering
        \includegraphics[width=1\columnwidth]{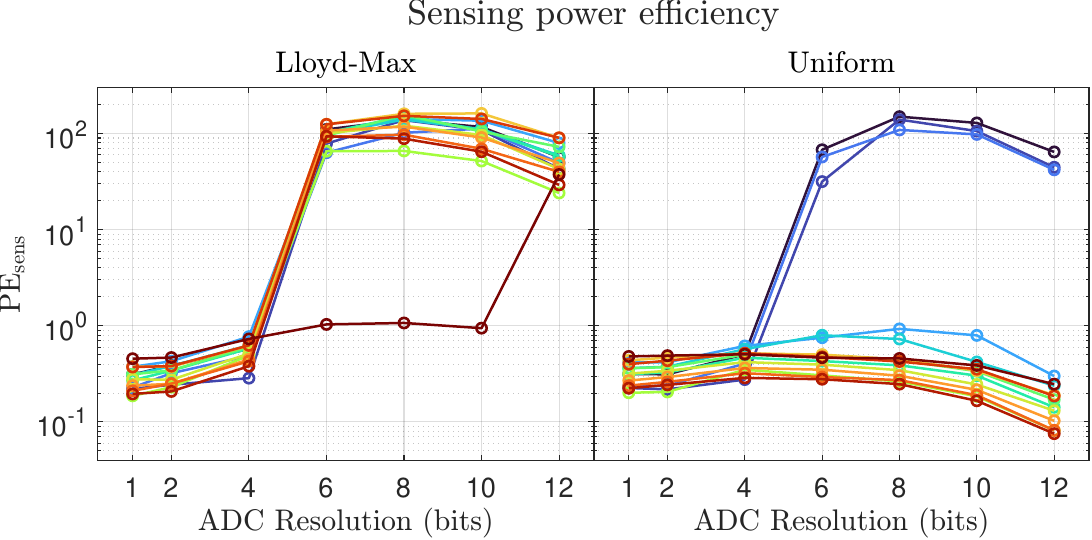}
        \caption{Sensing target estimation and power tradeoff.}
        \label{subfig:sens_pwr}
    \end{subfigure} \hfill
    \begin{subfigure}{1\columnwidth}
        \centering
        \includegraphics[width=1\columnwidth]{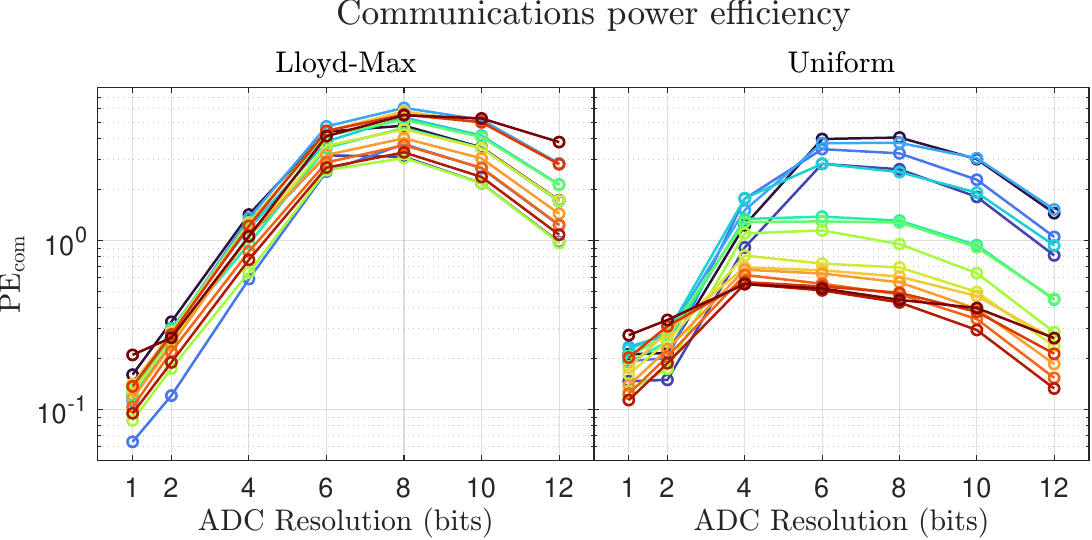}
        \caption{Communications EVM and power tradeoff.}
        \label{subfig:comms_pwr}
    \end{subfigure}
    \caption{Power consumption curves and the corresponding median channel estimation, sensing, and communications performance tradeoffs. The beamformer topologies in the legend are presented as $(D,N_a)$, a tuple containing the number of \ac{RF} chains and the subarray size. The order of the curves is sorted in descending order of best channel estimation performance with a uniform quantizer.}\label{fig:pwr_tradeoffs}
\end{figure*}

\vspace{-0.3cm}
\subsection{Power and Performance Tradeoffs}
Now we will study the power efficiency and performance tradeoffs of a larger set of beamforming topologies, containing fully connected, partially connected, and sub-panel topologies. We consider an identical scenario as in the previous simulations. We start by simply displaying the power consumption curves obtained by applying (\ref{eq:pwr_model}) to different numbers of \ac{RF} chains and \ac{ADC} resolutions. We present the resulting curves in Fig.~\ref{subfig:pwr_cons}. In this subsection, the beamformer topologies will be encoded as $(D,N_a)$ pairs. As explained in Section~\ref{subsec:perf_metrics}, we present the inverse products between the power consumption and the median relative channel estimation error, $L$th-PPE, and \ac{EVM}, in Figs.~\ref{subfig:ch_est_pwr}, \ref{subfig:sens_pwr}, and \ref{subfig:comms_pwr}, respectively. The order of the curves in the legend is sorted in descending order of best $\text{PE}_\text{ch}$ value for the uniform quantizer case.

Focusing on the \ac{LM} quantizer curves, it can be seen that the performance with optimal quantizers is generally similar for all topologies except $(1,8)$, the only topology with a single \ac{RF} chain. Additionally, there seems to be a trend maximum efficiency around \ac{ADC} resolutions of 6 to 8 bits, regardless of topology. Using higher resolution \ac{ADC} does improve performance, but seemingly not enough to justify the increase in power consumption.

%Regarding the performance with uniform quantizers, we now notice a larger performance gap between the best performing topologies---specifically $(4,2)$, $(8,1)$, and $(6,3)$---and the remainder. Curiously, one of the best performing topologies is the fully digital beamformer $(8,1)$ with 8 bits. Since quantization errors using uniform quantizers are naturally larger than when using \ac{LM} quantizers, this can indicate that these specific topologies are particularly robust to quantization error. Alternatively, this performance gap may be a consequence of the way the \ac{ADC} ends of scale are defined and the chosen value of \ac{Rx} backoff factor. Nevertheless, the best performing \ac{ADC} resolutions are still within the 6 to 8 bit range.

Regarding the performance with uniform quantizers, we now notice a larger performance gap between the best performing topologies---specifically $(4,2)$, $(8,1)$, and $(6,3)$---and the remainder. Curiously, one of the best performing topologies is the fully digital beamformer $(8,1)$ with 8 bits. \textcolor{blue}{This performance gap is a consequence of the interplay between the \ac{ADC} ends of scale (which is a function of the \ac{Rx} backoff factor) and a dominant \ac{LOS} path. Configurations with larger $N_a$ experience substantial clipping when the beam of a subarray is steered towards the \ac{LOS} \ac{AoA}. Meanwhile, the configurations with small $N_a$ observe much smaller amplitude variations at the input of the \ac{ADC} and, thus, do not clip as hard even at the same \ac{Rx} backoff, which explains the observed behavior in Fig.~\ref{fig:pwr_tradeoffs}. In general}, the best performing \ac{ADC} resolutions are still within the 6 to 8 bit range \textcolor{blue}{regardless of \ac{ADC} type.}

To represent the overall power efficiency characteristics of each topology, we combine the presented metrics into an \textit{ad hoc} average normalized metric. Let $\text{PE}_\text{x}(D,N_a,N^\text{bits}_\text{ADC})$ denote the power efficiency as a function of the number of \ac{RF} chains $D$, subarray size $N_a$, and \ac{ADC} resolution $N^\text{bits}_\text{ADC}$---where $\text{x}$ is either ``ch'' (channel estimation), ``sens'' (sensing), or ``com'' (communications)---then we first normalize the performance of each configuration by the maximum achieved value \vspace{-0.15cm}
\begin{equation}
    \overline{\text{PE}}_\text{x}(D,N_a,N^\text{bits}_\text{ADC}) = \frac{\text{PE}_\text{x}(D,N_a,N^\text{bits}_\text{ADC})}{\max_{(D,N_a,N^\text{bits}_\text{ADC})} \text{PE}_\text{x}(D,N_a,N^\text{bits}_\text{ADC})},
\end{equation}
where the maximum values are distinct for \ac{LM} and uniform quantizers.
We choose to average out the channel estimation, sensing, and communications performances, yielding the \ac{ANPE} metric
\begin{equation}
    \overline{\text{PE}}(D,N_a,N^\text{bits}_\text{ADC}) = \frac{\overline{\text{PE}}_\text{ch} + \overline{\text{PE}}_\text{sens} + \overline{\text{PE}}_\text{com}}{3},
\end{equation}
where we omitted the dependency on $(D,N_a,N^\text{bits}_\text{ADC})$. We consider this metric mostly for the purposes of data analysis and visualization, not intending for it to be considered as a performance metric for the study of power efficiency in more general contexts.

\begin{figure}[!htbp]
    \centering
    \includegraphics[width=1\linewidth]{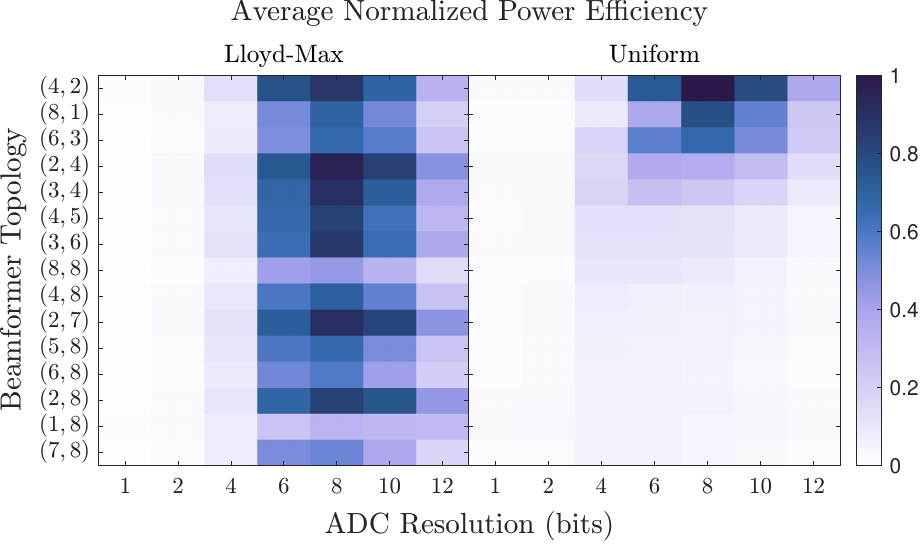}
    \caption{\ac{ANPE} values for the considered $(D,N_a)$ pairs and varying \ac{ADC} resolution. The $(D,N_a)$ pairs are displayed in descending order of best \ac{ANPE} value with a uniform quantizer.}
    \label{fig:normalized_pe}
\end{figure}

The collected \ac{ANPE} values are displayed in Fig.~\ref{fig:normalized_pe}, which visually summarizes some of the same conclusions previously reached in this subsection. Notably, with \ac{LM} quantizers, almost all topologies perform similarly well within the $N^\text{bits}_\text{ADC}\in\{ 6,8,10 \}$ range. For uniform quantizers, we once again see the top-3 topologies dominating the performance curves within a similar \ac{ADC} resolution range.

% % Another table showing the top 3 best performing configurations for each metric
% \begin{table*}[t]
%     \textcolor{blue}{
%     \centering
%     \begin{tabular}{|c||c|c|c||c|c|c|}
%          \cline{2-7}
%          \multicolumn{1}{c|}{} & \multicolumn{3}{c||}{Lloyd-Max} & \multicolumn{3}{c|}{Uniform}\\
%          \cline{2-7} \hline
%          Metric & Best | ADC & 2\textsuperscript{nd} Best | ADC  &  3\textsuperscript{rd} Best | ADC  & Best | ADC  & 2\textsuperscript{nd} Best | ADC  &  3\textsuperscript{rd} Best | ADC  \\
%          \hline
%          $\text{PE}^\star_\text{ch}$ & $(4,2)$ | 6 bits & $(3,4)$ | 8 bits & $(2,4)$ | 8 bits & $(4,2)$ | 8 bits & $(8,1)$ | 8 bits & $(6,3)$ | 6 bits \\
%          $\text{PE}^\star_\text{sens}$ & $(2,8)$ | 8 bits & $(2,7)$ | 8 bits & $(3,6)$ | 8 bits & $(8,1)$ | 8 bits & $(4,2)$ | 8 bits & $(6,3)$ | 8 bits \\
%          $\text{PE}^\star_\text{comms}$ & $(2,4)$ | 8 bits & $(2,7)$ | 8 bits & $(1,8)$ | 8 bits & $(4,2)$ | 8 bits & $(2,4)$ | 6 bits & $(6,3)$ | 6 bits \\
%          \hline
%     \end{tabular}
%   \caption{\textcolor{blue}{Top 3 beamformer topologies with their respective best \ac{ADC} resolution for each power efficiency metric.}}}
%   \label{tab:best_configs}
% \end{table*}

From the conducted explorations, we also draw more general conclusions, resulting in the following general recommendations:
\begin{enumerate}
    \item Using a single \ac{RF} chain is clearly the worst choice among the considered architectures, since it is very sensitive to hardware impairment effects even with high-resolution \acp{ADC}.
    \item The fully digital architecture with high-resolution \acp{ADC} leads to a prohibitively large power consumption (as is well known in the literature), which is not compensated for by sufficiently improved performance.
    \item The fully digital approach with medium resolution \acp{ADC} -- e.g., from 6 to 10 bits -- is also a viable choice, since the power consumption that results from higher \ac{ADC} resolution (when compared to the 1 or 2 bit \ac{ADC} case) leads to sufficient performance benefit and robustness to justify the costs. 
    \item A balance between the single \ac{RF} chain and the fully digital architectures appears to be the most resource-efficient approach. 
    \item Using more \ac{RF} chains with medium resolution converters seems to lead to the best tradeoff between channel estimation, sensing, and communications performance.
    \item Adding \ac{RF} chains leads to improved robustness, but with significant additional power consumption costs from \ac{PA} biasing and the \acp{ADC}, especially when using high-resolution \acp{ADC}.  
    \item Topologies with smaller subarrays generally display more robust, energy-efficient performance under the studied conditions.
\end{enumerate}

\section{Conclusion} \label{sec:conclusion}

% In this work we have studied the effects of hardware impairments on different beamforming architectures, focusing on sensing and communications performance. A set of numerical results has been provided based on the considered models for the channel, hardware impairments, and power consumption. These results explore the impact of the number of \ac{RF} chains and \ac{ADC} resolution on channel estimation, sensing, and communications performance. Then, a set of general suggestions aimed at orienting radio transceiver architecture choices for \ac{ISAC} is provided. In particular, we have shown that, in the studied scenario, fully analog and fully digital arrays do not provide efficient solutions when considering the performance and total cost tradeoff.\par

We have studied the effects of hardware impairments on different beamforming architectures, focusing on sensing and communications performance. This was done as an effort to understand how \ac{ISAC} can be achieved in an effective and robust manner with manageable total costs, which requires exploring hybrid beamforming solutions. \par

To enable our exploration on \ac{ISAC} with different beamformer topologies, we have introduced the concept of a \textit{doubly isotropic} pilot-combiner pair, formalizing the notion of an energy-fair beam sweep for \ac{OFDM} sensing. Additionally, based on intuition from previous work on channel estimation, we introduce the \ac{MS}-\ac{SAGE} algorithm to overcome the optimization issues arising from the nonconvexity of the likelihood function. We also derive the coordinate update expressions for \ac{SAGE} with hybrid beamformers. A set of numerical results has been provided based on the considered models for the channel, hardware impairments, and power consumption. These results explore the impact of the number of \ac{RF} chains, subarray size, and \ac{ADC} resolution on channel estimation, sensing, and communications performance. Then, general suggestions aimed at orienting radio transceiver architecture choices for \ac{ISAC} are provided. We have shown that, in the studied scenario, medium resolution \acp{ADC} and smaller subarray sizes provide the most efficient solutions when considering the tradeoff between performance and total cost.

For future work, considering a more realistic model of the amplifier nonlinearities, also including memory effects, seems like a prospective line of inquiry. Also, \ac{DPD} is a highly relevant element of practical \ac{RF} transmitters, thus, proper modeling of \ac{DPD} can also be included. On the signal processing aspect, addressing the problem of model order estimation with impaired hardware is a valid research direction. %can lead to a substantial contribution to the field of \ac{PCE}.

\vspace{-0.5cm}

\appendices

\section{Derivation of the Model Norm} \label{appendix:model_norm_deriv}
The model square norm is equivalent to
\begin{equation} \label{eq:model_norm_generic}
    \| \alpha(\boldsymbol{\xi}) \|^2_F =  \sum_{ntm} |r_{mt}(\psi) |^2 | x_{nt}(\varsigma) |^2 . 
\end{equation}
From the theory of isotropic pilots and from (\ref{eq:x_norm_expansion}) using orthogonal pilots we have 
\begin{multline}
    \sum_{nt} |x_{nt}(\varsigma)|^2 = \sum_{nt} \sum^{D-1}_{d=0} |\mathbf{a}^\text{T}(\varsigma) [\mathcal{P}]_{:dt} |^2 |s_{ntd}|^2 \\ = \sum^{K_\text{T}}_{k=1} |\mathbf{a}^\text{T}(\varsigma) [\mathcal{P}]_{:k} |^2 |s_{k}|^2, 
\end{multline}
where the $(n,t,d)$ tuples were grouped by precoder direction indexed by $k$ as in Section \ref{subsec:comb_pilot_analog}. We know that we require $|s_{k_1}| = |s_{k_2}| \forall k_1,k_2$ pairs for an isotropic pilot, thus \vspace{-0.2cm}
\begin{equation}
    \sum_{nt} |x_{nt}(\varsigma)|^2 =  \sum^K_{k=1} |\mathbf{a}^\text{T}(\varsigma) [\mathcal{P}]_{:k}|^2 |s_{k}|^2, 
    %\sum_{nt} |x_{nt}(\varsigma)|^2 = \frac{\rho}{N^\text{T}_a} \sum^K_{k=1} |\mathbf{a}^\text{T}(\varsigma) [\mathcal{P}]_{:k}|^2 |s_{k}|^2, 
\end{equation}
where $|s_{k}|^2 = \rho_k/N^\text{T}_a$ (since power is equally divided between all antennas in a subarray), $\rho_k$ is the total power available for each tuple $(n,t,d)$, which typically depends exclusively on the number of OFDM symbols allocated to each precoder direction. Here, denoting the total transmitted power by $\rho$, we have $\rho_k=\frac{\rho N^\text{T}_\text{S}}{K_\text{T}}$,i.e., the total energy delivered per precoder direction is equal to the total energy of a precoder subsequence (with length $N^\text{T}_\text{S}$) divided by the number of precoders. We can also express
\begin{equation} \label{eq:norm_prec_sqr}
    |\mathbf{a}^\text{T}(\varsigma)[\mathcal{P}]_{:k}|^2 = \!\!\!\!\! \sum_{v_1, v_2 \in \mathcal{V}_k} \!\!\!\!\!  e^{j(v_1 - v_2) \Delta\varsigma_k}\text{, with }\Delta\varsigma_k = \varsigma - \Bar{\varsigma}_k,
\end{equation}
where $\mathcal{V}_k$ denotes the contiguous set of antennas that are active for precoder $k$, it is implied that $|\mathcal{V}_k| = N^\text{T}_a$.
It is also not hard to show that \vspace{-0.3cm}
\begin{equation}
    |\mathbf{a}^\text{T}(\varsigma)[\mathcal{P}]_{:k}|^2  =  N^\text{T}_a + 2 \!\!\! \sum^{N^\text{T}_a-1}_{v=1} (N^\text{T}_a-v) \cos(v\Delta\varsigma_k), \label{eq:prec_asympt}
\end{equation}
and that, for $\frac{v}{K} \notin \mathbb{Z}$, \vspace{-0.3cm}
\begin{multline}
    \sum^{K-1}_{k=0} \cos\left(\frac{2\pi}{K_\text{T}}vk + v (\varsigma - \Bar{\varsigma}_0)\right)\\
    = \sin(\pi v) \csc\left( \frac{\pi v}{K_\text{T}} \right)
    \cdot \cos\left( \left(\frac{K_\text{T}-1}{K_\text{T}}\pi + \varsigma - \Bar{\varsigma}_0\right)v\right). \label{eq:sum_cos_K}
\end{multline}
Thus it is clear that, as long as $K_\text{T}\geq N^T_a$, then (\ref{eq:prec_asympt}) yields $\sum_k |\mathbf{a}^\text{T}(\varsigma)\mathbf{p}_{k}|^2  = K_\text{T} N^\text{T}_a$. Therefore $\sum_{nt} |x_{nt}(\varsigma)|^2 = \rho N^\text{T}_\text{S}$.

It remains now to compute $\sum_{(n,t,m)\in \mathcal{M}_k} |r_{mt} (\psi)|^2 $. We proceed similarly, by grouping tuples $(m,t)$ in sets $\mathcal{M}_k$ and choosing combiner directions similarly to what was done for the precoders. We then write \vspace{-0.2cm}
\begin{equation}
    \sum_{(n,t,m)\in \mathcal{M}_k}  \!\!\!\!\!\!\!\! |r_{mt} (\psi)|^2 = \sum^{K_\text{R}}_{k=1} |r_{k} (\psi)|^2,
\end{equation}
in which we assume that $K_\text{R} = D_\text{R} N^\text{R}_\text{S}$ and that there is no repetition of combiner directions. Similarly to (\ref{eq:norm_prec_sqr}), (\ref{eq:prec_asympt}), and (\ref{eq:sum_cos_K}), we have that
\begin{equation}
    \sum_{k} |r_{k} (\psi)|^2 = K_\text{R} N^\text{R}_a = D_\text{R} N^\text{R}_\text{S} N^\text{R}_a.
\end{equation}
Combining the results for the pilot and the combiners we get
\begin{equation}
    \sum_{ntm} |r_{mt}(\psi_\ell)|^2 |x_{nt}(\varsigma_\ell)|^2 = D_\text{R} N^\text{R}_a N_\text{S} \rho,
\end{equation}
where $N_\text{S} = N^\text{T}_\text{S} N^\text{R}_\text{S}$, i.e., the product of the sizes of the subsequences is equal to the total pilot length.

\section{Derivation of the Model Norm Beamforming Gain} \label{appendix:bf_gain}
Here we derive the model norm for the communications case, with matched combining and precoding towards the strongest path. From the model norm expression in (\ref{eq:model_norm_generic}), consider that 
\begin{equation}
    |x_{nt}|^2 = \frac{\rho}{N_\text{T}N_\text{C}} \left| \sum^{N_\text{T}-1}_{v=0} e^{-jv\varsigma_\ell} p_{vt} \right|^2 = \frac{\rho N_\text{T}}{N_\text{C}},
\end{equation}
where the last equality comes from the matched beamforming. Similarly, we have %\vspace{-0.2c}
\begin{equation}
    |r_{mt}|^2 = \left| \sum^{N_\text{R}-1}_{u=0} e^{-ju\psi_\ell} r_{mtu} \right|^2 = {N^\text{R}_a}^2,
\end{equation}
since $r_{mtu}$ is a matched combiner with $N^\text{R}_a$ nonzero elements due to the sub-panel topology. We then have that the matched beamforming model norm is
\begin{equation}
    \sum_{ntm} |x_{nt}|^2 |r_{mt}|^2 = D_\text{R} {N^\text{R}_a}^2 N_\text{T} N_\text{S} \rho,
\end{equation}
which, if divided by the doubly-isotropic model norm in (\ref{eq:model_norm}), yields a gain of $a_\text{comms} = N^\text{R}_a N_\text{T}$.

% \section{Derivation of the Computational Complexity?}

\vspace{-0.5cm}
\section*{Acknowledgment}
The authors would like to thank Visa Tapio and Harri Saarnisaari for the insightful suggestions and comments.

\ifCLASSOPTIONcaptionsoff
  \newpage
\fi



%\bibliographystyle{IEEEtran}

\bibliographystyle{IEEEtran}
\bibliography{IEEEabrv,biblio}
\end{document}